\documentclass[%
 reprint,
 amsmath,amssymb,
 aps,
prb,
]{revtex4-2}

\usepackage[utf8]{inputenc}
\usepackage{graphicx}
\usepackage{overpic}
 \usepackage{rotating}
 \usepackage{mwe, tikz}
 \usepackage{amsmath,amssymb}
\usepackage{pgfplots}
\usetikzlibrary{calc,arrows,decorations.markings,shapes.geometric} 
\usepackage{amssymb}

\usepackage[finalnew]{trackchanges}

\usepackage{lipsum}
\usepackage{romannum}
\usepackage{xfp}

\soulregister\ref7
\soulregister\cite7
\soulregister\citeA7
\soulregister\romannum7

\usepackage{graphicx}
\usepackage{amsmath}
\usepackage{mathtools} 
\usepackage{color}
\usepackage{overpic}
\usepackage[export]{adjustbox}
\usepackage{mwe, tikz}
\usepackage{rotating}
\usepackage{latexsym}
\usepackage{csquotes}
\usepackage{tikz} 
\usetikzlibrary{calc,arrows,decorations.markings} 
\usepackage{pict2e}
\usepackage{pgfplots}
\usepackage{float}
\usepackage{hyperref}
\usepackage{xr}
\usepackage{rotating}
\usepackage{amsmath,amssymb}
\usepackage{fp}
\usetikzlibrary{calc} 
\usepackage{booktabs, tabularx}
\usepackage{romannum}
\usepackage{bm}
\usepackage{float}
\usepackage{adjustbox}
\usepackage{booktabs, tabularx}
\usepackage{xfp}
\usepackage{ifthen}
\usepackage{pgffor}
\usepackage[caption=false]{subfig}
\newcommand{\legSqTxt}[3]{
		\begin{tikzpicture}
			     \coordinate (center1) at (#1,#2); 
                \coordinate (b) at ($ (center1) - (.075,0) $);
                \coordinate (c) at ($ (b) - (.1,0) $); 
                \coordinate (d) at ($ (center1) + (.075,0) $);
                \coordinate (e) at ($ (d) + (.1,0) $); 
				 \draw[#3,fill=#3] ($(b)-(0.0,0.075)$) rectangle ($ (d) + (0.0,0.075) $); 
		\end{tikzpicture}		
}

\newcommand{\circTxtFill}[3]{
		\begin{tikzpicture}
                \coordinate (center1) at (#1,#2); 
				 \draw[#3,fill=#3] (center1) circle (2pt);
		\end{tikzpicture}		
}

\definecolor{light-gray}{gray}{0.85}

\newcommand{\drawSlope}[6]{ 
				\coordinate (center1) at (#1,#2); 
				 \FPeval{\nx}{cos(#4*pi/180)}%
				 \FPeval{\ny}{sin(#4*pi/180)}%
				 \FPeval{\absNx}{abs(\nx)}%
				 \FPeval{\absNy}{abs(\ny)}%
				\coordinate (b) at ($ (center1) + #3*(-\ny,+\nx) $);
				\coordinate (c) at ($ (center1) - #3*(-\ny,+\nx) $); 
				 \FPeval{\xx}{(-\nx*\ny)}%
				\ifdim\xx pt < 0pt 
				\coordinate (d) at ($ (c) -2*#3*\ny*(1,0) $);
				\node at ($(d) +(0,#3*\nx)-0.2*(\nx/\absNx,0)$) {{\color{#5}#6}}; 
				\else
				\coordinate (d) at ($ (c) +2*#3*\nx*(0,1) $);
				\node at ($(d)-#3*\nx*(0,1)-0.2*\nx/\absNx*(1,0)$) {{\color{#5}#6}}; 
				\fi
				\draw[#5,line width=0.1mm] (d) -- (b); 
				\draw[#5,line width=0.1mm] (d) -- (c); 
				\draw[#5,line width=0.1mm] (b) -- (c); 
} 

\newcommand*{\Labelxy}[4]{\put(#1,#2) {\setlength{\fboxsep}{0pt}{\strut\textcolor{black}{\begin{turn}{#3}{#4}\end{turn}}}}}
\newcommand*{\LabelFig}[3]{\put(#1,#2) {\setlength{\fboxsep}{0pt}\colorbox{white}{\textcolor{black}{#3}}} }
\definecolor{darkolivegreen}{rgb}{0.33, 0.42, 0.18}
\definecolor{darkspringgreen}{rgb}{0.09, 0.45, 0.27}
\definecolor{darkslategray}{rgb}{0.18, 0.31, 0.31}
\definecolor{darkred}{rgb}{0.55, 0.0, 0.0}
\addeditor{KK}
\usepackage{xr}
\makeatletter
\newcommand*{\addFileDependency}[1]{
  \typeout{(#1)}
  \@addtofilelist{#1}
  \IfFileExists{#1}{}{\typeout{No file #1.}}
}
\makeatother
\newcommand*{\myexternaldocument}[1]{
    \externaldocument{#1}
    \addFileDependency{#1.tex}
    \addFileDependency{#1.aux}
}
\myexternaldocument{./sm}

\begin{document}

\title{Prediction of steel nanohardness by using graph neural networks on surface polycrystallinity maps}

\author{Kamran Karimi$^1$}
\email{kamran.karimi@ncbj.gov.pl}
\author{Henri Salmenjoki$^2$}
\author{Katarzyna Mulewska$^1$}
\author{Lukasz Kurpaska$^1$}
\author{Anna Kosińska$^1$}
\author{Mikko J. Alava$^{1,2}$}
\author{Stefanos Papanikolaou$^1$}
\email{stefanos.papanikolaou@ncbj.gov.pl}
\affiliation{%
 $^1$ NOMATEN Centre of Excellence, National Center for Nuclear Research, ul. A. Sołtana 7, 05-400 Swierk/Otwock, Poland\\
 $^{2}$ Aalto University, Department of Applied Physics, PO Box 11000, 00076 Aalto, Espoo, Finland
}%

\begin{abstract}
As a bulk mechanical property, nanoscale hardness in polycrystalline metals is strongly dependent on microstructural features that are believed to be heavily influenced from complex features of polycrystallinity ---namely, individual  grain orientations and neighboring grain properties.
We train a graph neural network (GNN) model, with each grain center location being a graph node, to assess the predictability of micromechanical responses of nano-indented low-carbon 310S stainless steel (alloyed with Ni and Cr) surfaces, solely based on surface polycrystallinity, captured by electron backscatter diffraction maps. 
The grain size distribution ranges between $1-100~\mu$m, with mean grain size at $18~\mu$m.
The GNN model is trained on a set of nanomechanical load-displacement curves, obtained from nanoindentation tests  and is subsequently used to make predictions of nano-hardness at various depths, with sole input being the grain locations and orientations. 
Model training is based on a sub-standard set of $\sim10^2$ hardness measurements, leading to an overall satisfactory performance. 
We explore  model performance and its dependence on various structural/topological grain-level descriptors, such as the grain size and number of nearest neighbors.
Analogous GNN model frameworks may be utilized for quick, inexpensive hardness estimates, for guidance to detailed nanoindentation experiments, akin to cartography tool developments in the world exploration era. \note[KK]{WHAT IS THIS? it's just an analogy; non-indented grains depicted as hatched areas in Fig.~\ref{fig:hmap} closely resemble unmapped/unknown geographical regions in early cartography maps!} 

\end{abstract}

\maketitle

\section{Introduction}
Polycrystals consist of complex crystalline grain networks that are known to dictate {multiscale mechanical} responses \cite{wadhwa2008textbook}. 
Nevertheless, inherent microstructure-property-process correlations may not be typically captured by constitutive relations and contain overwhelming complexity~\cite{lasalmonie1986influence}. 
A remarkable exception is the famous Hall–Petch relationship \cite{hall1951deformation,petch1953cleavage}, connecting grain size and strength \cite{alavudeen2006textbook,yang2005grain}.
Primary strengthening mechanisms, such as dislocation pile-ups and slip transfer capacity (across adjacent grains), are closely tied to intrinsic geometry of grains as well as their crystallographic orientation and associated degree of misalignment across boundaries \cite{wang2004indentation,britton2009nanoindentation,pathak2012studying,ohmura2005nanoindentation}.
Indeed, conventional phenomenological frameworks are limited in these respects, thus significantly restricting their predictive capacities~\cite{wadhwa2008textbook}. 
In this paper, we construct a machine-learned graph neural network (GNN)-based framework from a fairly large ensemble of intrinsic structural features associated with the complex polycrystallinity of 310S steel.
Using a relevant set of nano-mechanical  tests, our supervised model is trained to produce interpretable predictions of micromechanical responses and indentation hardness solely based on (an appropriate suite of) microstructural predictors.
The proposed framework may complement elaborate experimental and numerical investigations of metals' surface polycrystallinity by drastically improving material surface exploration for mechanical purposes.   

\begin{figure*}[t]
    \centering
    \begin{overpic}[width=\textwidth]{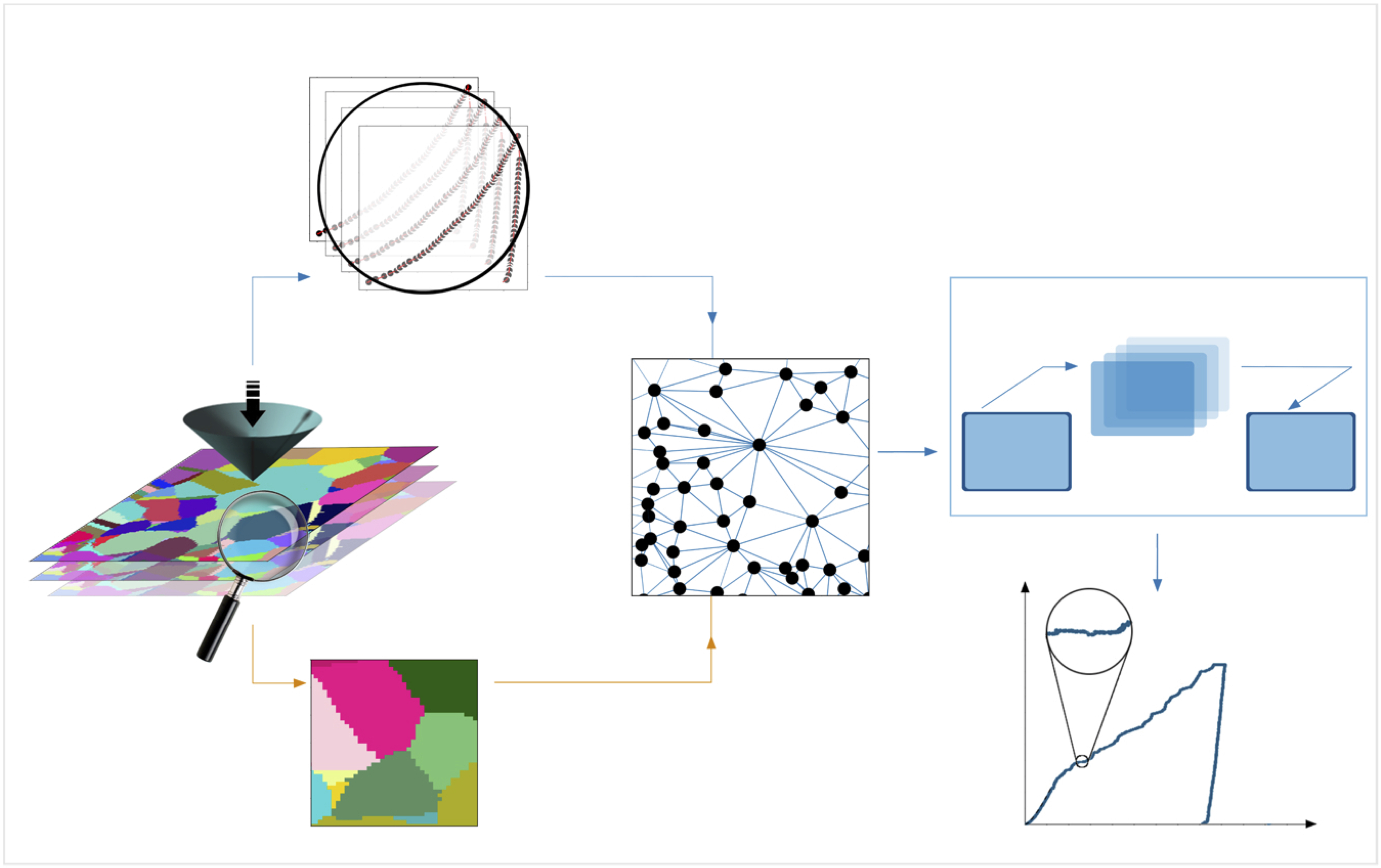}
        %
        \Labelxy{29}{40}{0}{Depth}
        \Labelxy{20}{50}{90}{Load}
        \Labelxy{43}{40}{0}{\color{cyan} Target}
        \Labelxy{1}{23}{35}{Nanoindentation}
        \Labelxy{42}{21}{90}{Graph Neural Net}
        \Labelxy{71}{29.5}{0}{Encode} %
        \Labelxy{81}{33.5}{0}{Core}
        \Labelxy{91.5}{29.5}{0}{Decode}
        \Labelxy{77}{37}{45}{$\scriptstyle\times n_\text{proc}$}
        \Labelxy{70}{44}{0}{Encode-Process-Decode Architecture}
        \Labelxy{27}{11}{0}{{$i$}}
        \Labelxy{32}{8}{0}{{$j$}}
        \Labelxy{35}{9.5}{0}{Nodal attributes $ {v}_i:$}
        \Labelxy{35}{8}{0}{\scriptsize (grain size, perimeter, \# of neighbors, ...)}
        \Labelxy{35}{5}{0}{Edge attributes $ {e}_{ij}:$}
        \Labelxy{35}{3.4}{0}{\scriptsize (misorientation angle)}
        \Labelxy{42}{14.5}{0}{\color{orange} Predictors}
        \Labelxy{93}{1}{0}{Depth}
        \Labelxy{72}{18}{90}{Load}
        \Labelxy{83}{17}{0}{Pop-in}
        \Labelxy{56}{17}{0}{{\large $\{ {v}_i\}_{i=1...n_g}$}}
        \Labelxy{56}{14}{0}{{\large $\{ {e}_{ij}\}_{i,j<i}$}}
        \begin{tikzpicture}
            \coordinate (a) at (0,0);
            \node[white] at (a) {\tiny.};                 
            \draw[cyan, line width=0.4mm] (9.5,2.5) -- (9.8,2.5);
            \draw[black,fill=black] (9.7,3) circle (.07cm);
            \drawSlope{15.85}{1.8}{0.7}{174}{black}{\hspace{1pt} \large $\frac{dp}{dh}$}
        \end{tikzpicture}
     \end{overpic}
    \caption{The supervised machine-learning workflow combining nanoindentation and scanning electron microscope EBSD experimental data with graph neural net (GNN) based approach. The load-depth curves obtained from a nanoindented specimen serve as the \emph{target} data for the GNN model. The grain microstructure and associated attributes are utilized as relevant \emph{predictors}. The combined data set is employed to construct a GNN, with each grain center location being a graph node, to be trained through the ``encode-process-decode" architecture. The model can learn from underlying correlations between the metallic microstructure and materials' mechanical response and make predictions on the latter (i.e. grain hardness) purely based on the former information. \note[KK]{IN THE FIGURE THE GNN PICTURE IS STRANGE. WHAT ARE THOSE GREY NODES?fixed.}}
    \label{fig:workFlow}
\end{figure*}
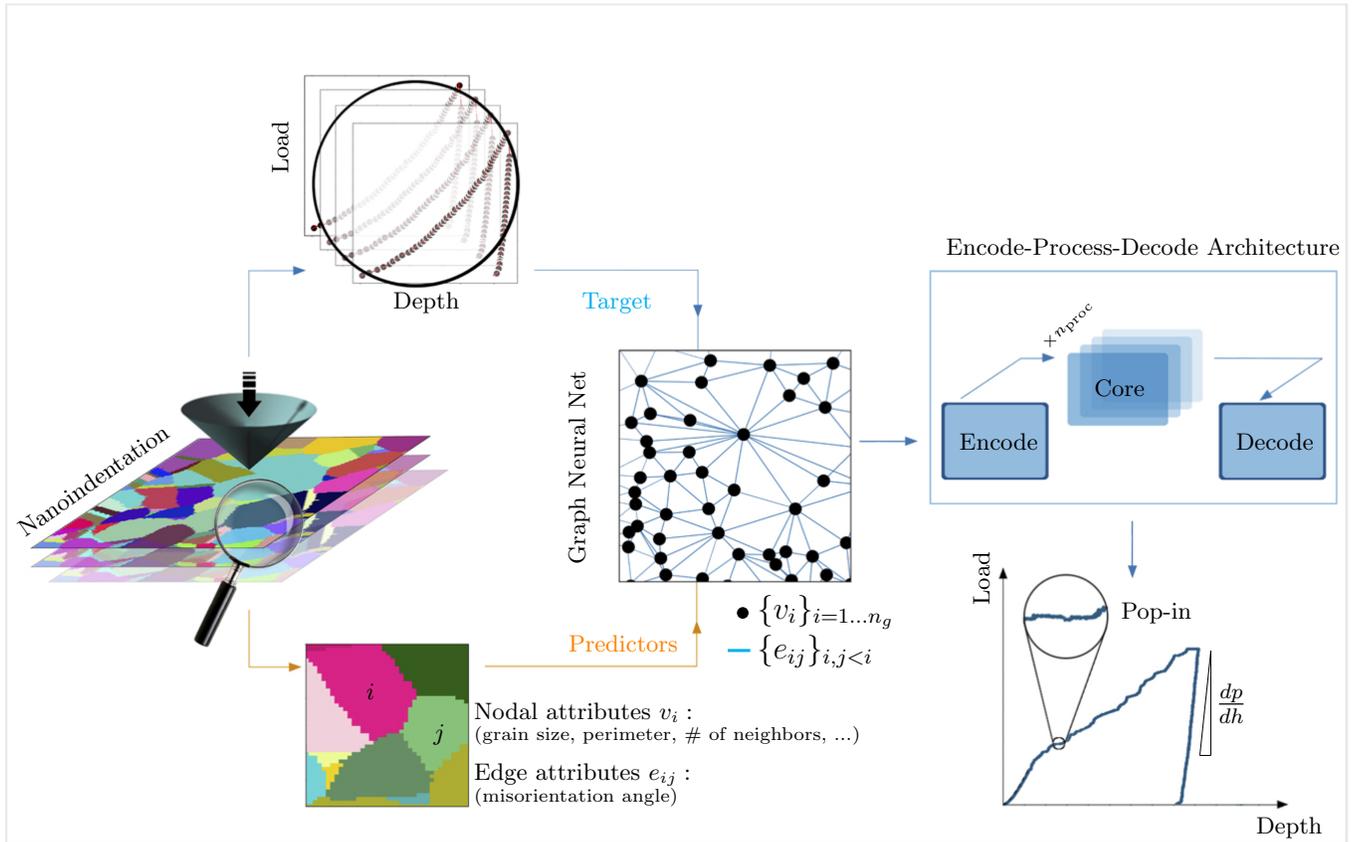
 
 Graph-based representations of polycrystals have been quite common in the attempt to describe microstructre-property relationships~\cite{grain-1,grain-2,grain-3,grain-4}. However, individual grain behaviors in polycrystals have been challenging to identify, with a wealth of constitutive parameters being commonly used to model them~\cite{grain-4}. GNNs provide a way to capture and learn these behaviors in a consistent way, that can then be used to predict mechanical responses, solely based on the grain environment. GNNs combine conveniences of both conventional (feature-based) machine learning methods and deep learning, but with unstructured architectures that are more adherent to real physical contexts \cite{battaglia2018relational,bronstein2017geometric,zhou2020graph, frydrych2021materials}.
GNN has been used in recent applications in materials science relevant to dynamics of glassy systems \cite{bapst2020unveiling,shiba2022unraveling} as well as property predictions in crystalline materials \cite{xie2018crystal} and some aspects of polycrystalline metals~\cite{pagan2022graph}.

\begin{figure}
    \centering
    %
    \begin{overpic}[width=0.28\textwidth]{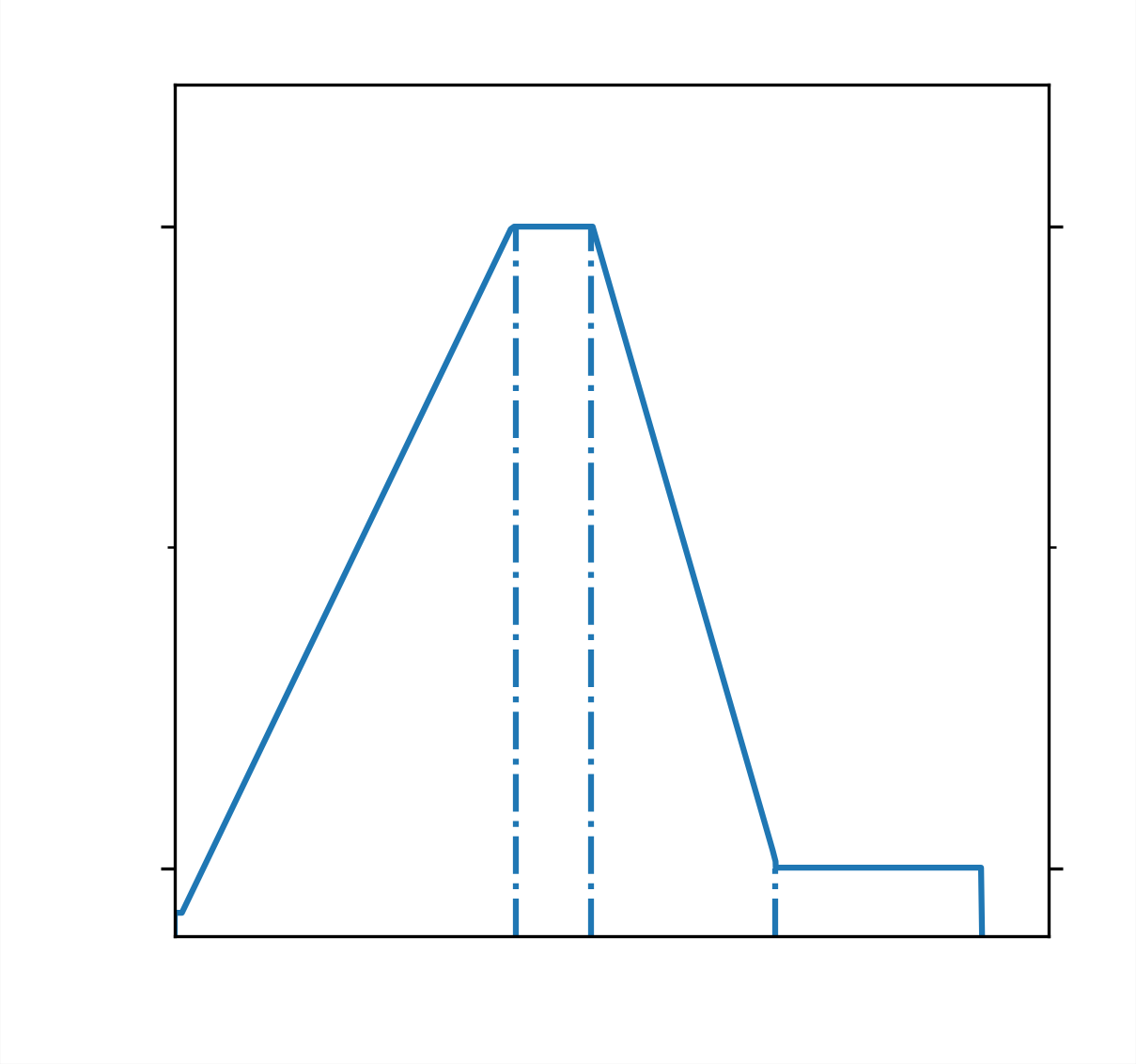}
        \Labelxy{4}{70}{90}{$f_\text{max}$}
        \Labelxy{4}{12}{90}{$f_\text{min}$}
        \Labelxy{26}{4}{0}{$t_\text{load}$}
        \Labelxy{40}{78}{0}{$t^0_\text{dwell}$}
        \Labelxy{53}{4}{0}{$t_\text{unload}$}
        \Labelxy{73}{4}{0}{$t^1_\text{dwell}$}
        \LabelFig{17}{79}{$a)$}
    \end{overpic}
    \begin{overpic}[width=0.28\textwidth]{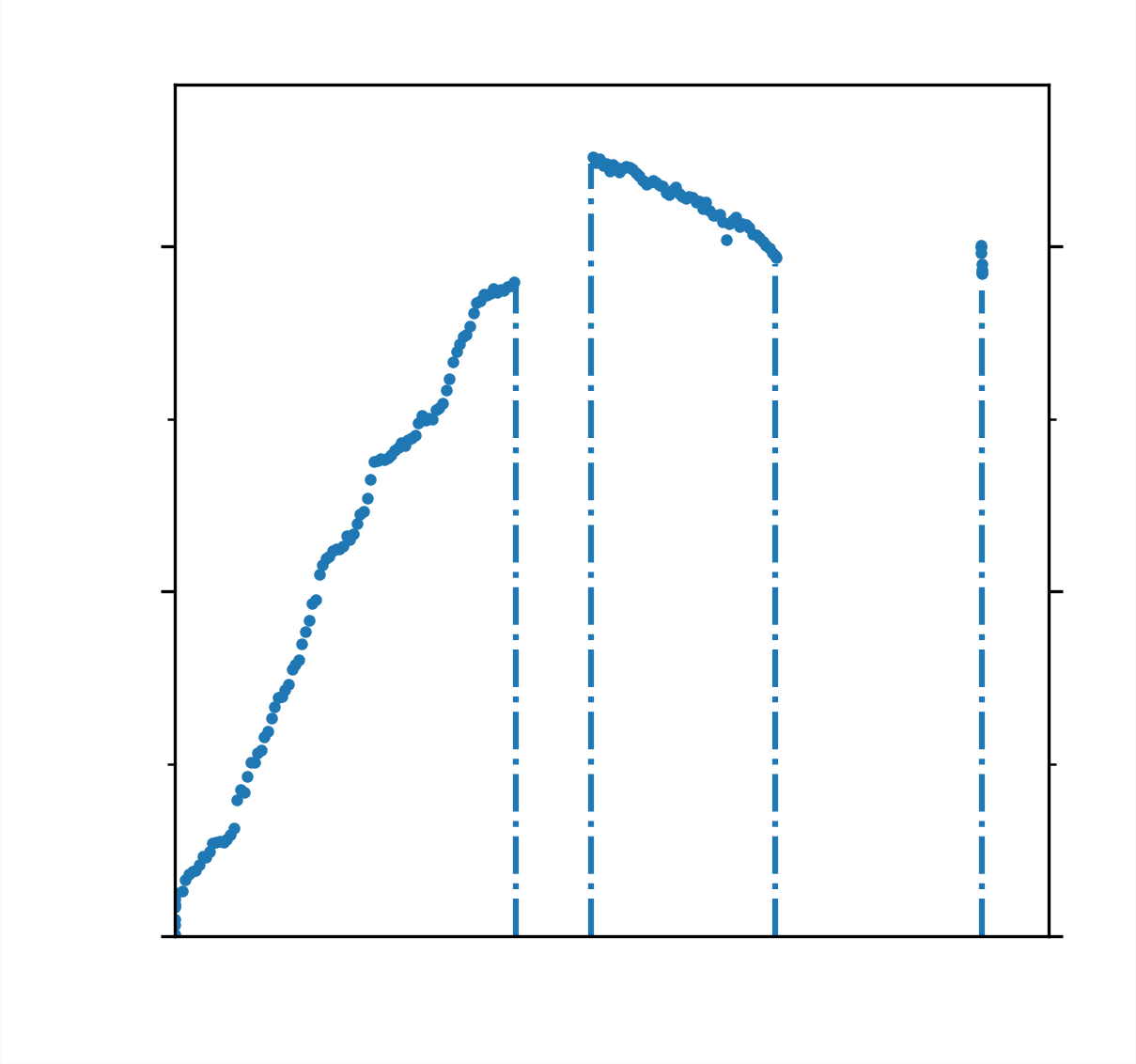}
        \Labelxy{4}{40}{90}{Depth}
        \Labelxy{26}{4}{0}{$t_\text{load}$}
        \Labelxy{53}{4}{0}{$t_\text{unload}$}
        \Labelxy{73}{4}{0}{$t^1_\text{dwell}$}
        \LabelFig{17}{79}{$b)$}
    \end{overpic}
    %
    \caption{Load-depth curves corresponding to the force-controlled nanoindentation tests.  {a}) dynamics of the applied force  {b}) measured indentation depths versus elapsed time. The graphs are not scale.}
    \label{fig:force_time}
\end{figure}

Nanoindentation tests provide valuable insights into complex microstructural strengthening and hardening mechanisms at the nanoscale, albeit with size effects~\cite{nix1998indentation} that mask bulk microstructural responses~\cite{bolin2019bending,durst2006indentation,papanikolaou2017avalanches}. Electron microscopy has been shown to significantly assist the interpretation of nanoindentation results in a wealth of materials~\cite{tem-1,tem-2,tem-3}.
Here, we integrate nanoindentation data, informed by electron backscatter diffraction (EBSD) mapping, with a supervised data-driven approach based on the graph neural net model to infer {micromechanical responses and} grain-scale hardness from the {surface polycrystallinity} information (see Fig.~\ref{fig:workFlow}).
We construct and train the GNN using an EBSD orientation imaging map containing {individual grains' orientations and neighboring grain properties} which was supplemented by a micromechanical data set corresponding to a nanoindented {low-carbon 310S stainless steel (alloyed with Ni and Cr)}.
The latter information consists of grain-scale load-displacement curves to be used as \emph{response} variables for the prediction task. Microstructural \emph{predictors} include an exhaustive list of (quantitative and categorical) grain-related characteristics with quite different scales that are all inferred and post-processed from the EBSD grain map (see Fig.~\ref{fig:EBSD}). \note[KK]{WHAT ARE THESE CHARACTERISTICS? SAY HERE. done.}
\add[KK]{This includes grain size (\texttt{area}), \texttt{perimeter}, length of inner boundaries (\texttt{subBoundaryLength}), \texttt{diameter}, perimeter of a circle with the same area (\texttt{equivalentPerimeter}), perimeter divided by equivalent perimeter (\texttt{shapeFactor}), a boundary grain (\texttt{isBoundary}), a grain with inclusions (\texttt{hasHole}), an inclusion grain (\texttt{isInclusion}), and number of neighboring grains (\texttt{numNeighbors}) as well as the misorientation angle (\texttt{misOrientationAngle}) between neighboring grains and associated boundary length (\texttt{boundaryLength}).}
The grains' (numerical) descriptors are typically distributed over a broad range of scales with large variations in the associated nanomechanical response.
The grain size distribution, as an example, has a lower cutoff at about $1~\mu$m and mean value of $18~\mu$m but is largely skewed with a long tail that extends up to $100$ microns.
Nevertheless, the overall predictive accuracy of our GNN model is fair given a relatively limited size of statistics (less than $200$ sample points).
This is verified in a systematic way by considering the learning process and its dependence on the training size as well as descriptor sets of varying size. 


 On top of high predictive accuracy, GNNs provide highly interpretable results and qualitative insights about underlying correlations between structural metrics and predicted nanomechanical response.  
More specifically, we find the grain diameter as a relevant hardness predictor which is in agreement with physics principles and could be verified in the context of the {well-established} Hall-Petch relationship, connecting the former and polycrystalline metals' strength (and/or hardness).
\change[KK]{In this framework, machine-learned models may improve the hardness \emph{exploration} within a multi-resolution hardness cartography map in polycrystalline metals which is otherwise impractical solely based on microscale nanoindentation tests.}{In this framework, machine-learned models may accelerate experimental investigations relevant to the hardness \emph{exploration} in polycrystalline metals.} \note[KK]{I DONT GET THIS SENTENCE. sentence reformulate.}
In addition to bulk mechanical properties (i.e. indentation hardness), our model may also be fine-tuned to accurately forecast indentation-induced strain bursts (i.e. pop-ins) \cite{kossman2021pop} and associated statistical distributions solely based on microstructural inputs.

\section{Methods}
\subsection{Nanoindentation Testing}
The sample preparation, nanoindentation experiments, and microstructural characterization of the low-carbon $310$S stainless steel but high in Ni ($19-22\%$) and Cr ($24-26\%$) content are detailed in \cite{dominguez2022mechanisms}.
We performed nanoindentation tests using the NanoTest Vantage system designed by Micro Materials Ltd. 
Hardness measurements were made at room temperature by using a Berkovich diamond indenter tip in a load-controlled manner at various depths.
Dynamics of the applied force is given in Fig.~\ref{fig:force_time}(a) with the maximum load level, denoted by $f_\text{max}$, exerted over the loading period of duration $t_\text{load}$ and following (first) dwell period $t^0_\text{dwell}$.
The specimen is subsequently unloaded to a residual force $f_\text{min}$ over the time scale $t_\text{unload}$ before it goes through the second dwell period of duration $t^1_\text{dwell}$ for thermal drift corrections.
The experiments were repeated over $12$ distinct $f_\text{max}$ values selected between $0.25-10$ mN and $15$ different indentation points per $f_\text{max}$ which were chosen to be about $20~\mu $m (the mean grain size) apart in distance .
This led to $180$ mechanical tests in total.
Here $f_\text{min}=0.25$ mN and $t^1_\text{dwell}=60$ s.
We also set $t_\text{load}=10$ s, $t^0_\text{dwell}=2$ s, $t_\text{unload}=5$ s for $f_\text{max}>5$ mN and $t_\text{load}=5$ s, $t^0_\text{dwell}=1$ s, $t_\text{unload}=3$ s otherwise.
As the outputs, we measure indentation depths as a function of time with a temporal resolution of order $\Delta t \simeq 0.05$ s, as in Fig.~\ref{fig:force_time}(b).

\subsection{Microstructural Characterization}
The microstructural characterization and EBSD analysis of the indented sample was performed through a ThermoFisher Scientific Helios 5 UX scanning electron microscope equipped with an EDAX Velocity Pro EBSD system. 
The grain mapping was performed using a $20$ keV electron beam with a $6.4$ nA probe current. 
The EBSD map was subsequently reconstructed through an EDAX OIM Analysis 8 software by grouping sets of (at least $2$) connected and similarly-oriented points (within $\pm 5^{\circ}$ uncertainties in angle) into individual grains. 
Crystallographic orientations, expressed in terms of Miller indices, can be assigned to each reconstructed grain (Fig.~\ref{fig:EBSD}) to be used as model inputs for the GNN framework.

\subsection{Training workflow}
\begin{figure}[t]
    \centering
    \begin{minipage}{0.3\textwidth} 
        \includegraphics[width=\linewidth]{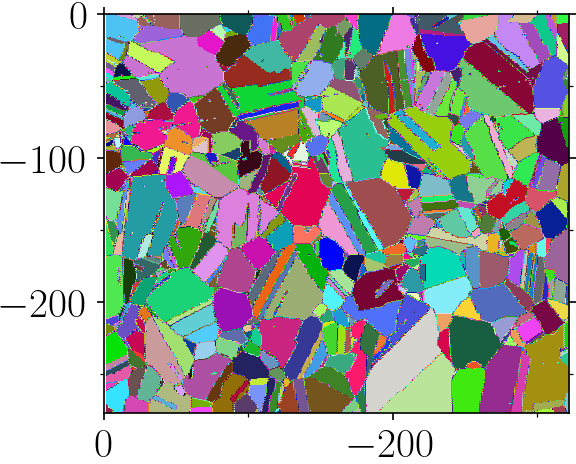}
         \Labelxy{-70}{-8}{0}{$x(\mu \text{m})$}
         \Labelxy{-166}{60}{90}{$y(\mu \text{m})$}
    \end{minipage}

    \bigskip
    \begin{minipage}{\columnwidth} 
        \centering
        \renewcommand\tabularxcolumn[1]{m{#1}}
        \renewcommand\arraystretch{1.3}
        \setlength\tabcolsep{12pt}
    \begin{tabularx}{\linewidth}{|p{2.9cm}|p{1.5cm}|p{1.3cm}} 
\hline\hline
\begin{tabular}{@{}l@{}} {microstructural} \\  {predictors}  \end{tabular} &
\begin{tabular}{@{}l@{}} {statistical} \\  {mean} \end{tabular} &
\begin{tabular}{@{}l@{}}\small {standard} \\  {deviation} \end{tabular} \\[2ex]
\hline\hline  
\texttt{x}  & $-120.9(\mu \text{m})$ & $75.5$  \\
\texttt{y}  & $-148.1(\mu \text{m})$ & $95.7$  \\
\texttt{area}   & $82.9(\mu \text{m}^2)$ & $192.5$ \\
\texttt{perimeter}   & $35.3(\mu \text{m})$ & $48.5$ \\
\texttt{subBoundaryLength}  & $0.1(\mu \text{m})$ & $0.9$ \\
\texttt{diameter}  & $11.0(\mu \text{m})$ & $13.7$ \\
\texttt{equivalentPerimeter}   & $20.1(\mu \text{m})$ & $25.2$ \\
\texttt{shapeFactor}  & $1.5$  & $0.5$ \\
\texttt{isBoundary}$^*$ & $0.1$ & $0.3$ \\
\texttt{hasHole}$^*$  & $0.1$ & $0.2$ \\
\texttt{isInclusion}$^*$  & $0.2$ & $0.4$ \\
\texttt{numNeighbors}  & $4.8$ & $4.3$ \\
\texttt{misOrientationAngle}$^\dagger$  & $45.0^\circ$ & $12.1$ \\
\texttt{boundaryLength}$^\dagger$
  & $7.3$ & $10.9$ \\
    \hline
    \end{tabularx}
         \Labelxy{-246}{-120}{0}{$^*~$\scriptsize Binary variables}
         \Labelxy{-246}{-128}{0}{$^\dagger~$\scriptsize Edge-based attributes}
        \vspace{+0.0pt}
    \end{minipage}
\caption{The EBSD map used to extract numerical and binary microstructural variables as the model input.}
    \label{fig:EBSD}
\end{figure}

Nanoindentation load-depth curves were obtained from tests performed independently on $n_\text{ind} = 131$ individual grains (out of $n_g=1080$ grains) based on a load-controlled protocol described above.
We discretized the corresponding displacement and force data (excluding the second dwell period) as a function of time into regular arrays of size $n_\text{dis}=100$, as in Fig.~\ref{fig:prediction_test} and \ref{fig:prediction_train}, and assembled the former in the target matrix $Y_{n_\text{ind}\times n_\text{dis}}$ to serve as training and test examples for our network.
For the case of (large) grains with multiple indentation sites, we simply used the average  displacement curves and our network was trained to predict deformation for individual grains.
To reconstruct the network from the EBSD map (Fig.~\ref{fig:EBSD}), each grain is treated as a separate node with index $i$ and $i=1...n_g$ in our graph. 
The graph connectivity is based on  neighboring grains; that is, grain index $i$ and $j$ sharing a common border on the grain map are connected by an edge $ij$ (see Fig.~\ref{fig:workFlow}). 

As a nodal \emph{feature}, every node is assigned two dimensional Cartesian coordinates (\texttt{x,y}) associated with the center of each grain.  
Additional nodal attributes extracted from the original map (and expected to correlate with the mechanical response) include grain size (\texttt{area}), \texttt{perimeter}, length of inner boundaries (\texttt{subBoundaryLength}), \texttt{diameter}, perimeter of a circle with the same area (\texttt{equivalentPerimeter}), perimeter divided by equivalent perimeter (\texttt{shapeFactor}), a boundary grain (\texttt{isBoundary}), a grain with inclusions (\texttt{hasHole}), an inclusion grain (\texttt{isInclusion}), and number of neighboring grains (\texttt{numNeighbors}).
The discretized force vector (as a control parameter in the experiment) was concatenated with the above set of structural attributes  of dimension $n_f=12$ with the assembled feature matrix given as $X_{n_\text{ind}\times (n_\text{dis}+n_f)}$.
To avoid features with significant variations in scale, every column of the above matrix was $z$-scored independently to have a zero mean and unit variance.
Furthermore, the edges of the graph accommodate the misorientation angle (\texttt{misOrientationAngle}) between two neighboring grains $i$ and $j$ along with the associated boundary length (\texttt{boundaryLength}) as their features \cite{bachmann2010inferential} .

\newcommand\blfootnote[1]{%
  \begingroup
  \renewcommand\thefootnote{}\footnote{#1}%
  \addtocounter{footnote}{-1}%
  \endgroup
}

The full set of nodal and edge-based features, i.e.  $\{ {v}_i\}_{i=1...n_g}$ and $\{ {e}_{ij}\}_{i,j<i}$ in the input graph, is initially encoded by an ``Encoder" block and subsequently processed via a ``Core" structure with $n_\text{proc}=3$ rounds of processing based on the message-passing framework \cite{battaglia2018relational}.
The ``Decoder" block returns an output graph (with the exact same structure as the input one) based on the Core's outcome but with predicted attributes, i.e. expected displacements, based on the nodal and edge-based descriptors.
Within the message-passing framework, the GNN applies two learning multilayer perceptrons (mlp) including an edge-based $\phi_e$ and node-based $\phi_v$ to each edge and node in order to compute updated node and edge attributes iteratively.
The two mlp's have identical architectures and are composed of two hidden layers and eight neurons per layer with a tangent hyperbolic activation function.

The optimization of the GNN model was performed by minimizing the loss function based on the mean-squared error (MSE) between the actual displacements and those outputted by the GNN using the stochastic gradient descent over the entire parameter space with a
learning rate of $10^{-3}$.
The graph data was split into training and testing sets and further trained using a four-fold cross validation.
We note that both sets (the training and testing examples) are present within the same graph and that the node labels associated with the test data (i.e. displacements) are invisible during the training process.
We use the GNNs library in Python which is DeepMind's implementation of graph neural nets based on Google's Tensorflow \cite{battaglia2018relational}. 

\section{Results}
We first investigate the GNN model and assess its predictive power of nanoindentation responses based on the grains' surface structure and given history of applied forces.
This includes a systematic analysis of the learning process of GNN from different subsets of existing grain-level predictors and varying training sizes.
As a further validation, we extract grain nano hardness at various depths from the predicted response and quantify how well the predictions compare with the actual data.

\subsection{Training and validation}
\begin{figure}
    \begin{overpic}[width=0.23\textwidth]{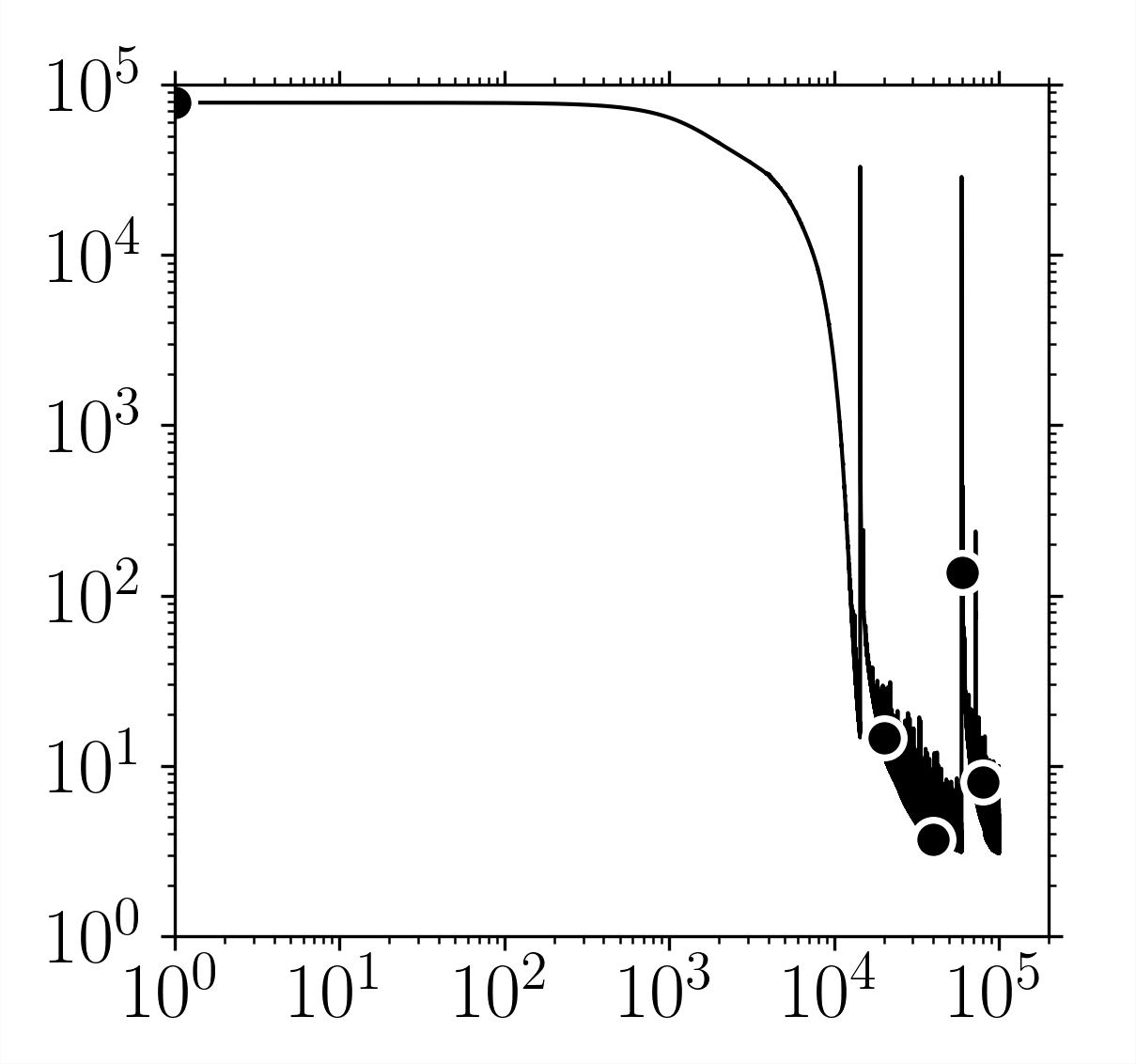}
        \LabelFig{17}{14}{$a)~\text{Train}$}
        \Labelxy{50}{-3}{0}{Epoch}
        \Labelxy{-3}{46}{90}{mse}
    \end{overpic}
    \begin{overpic}[width=0.23\textwidth]{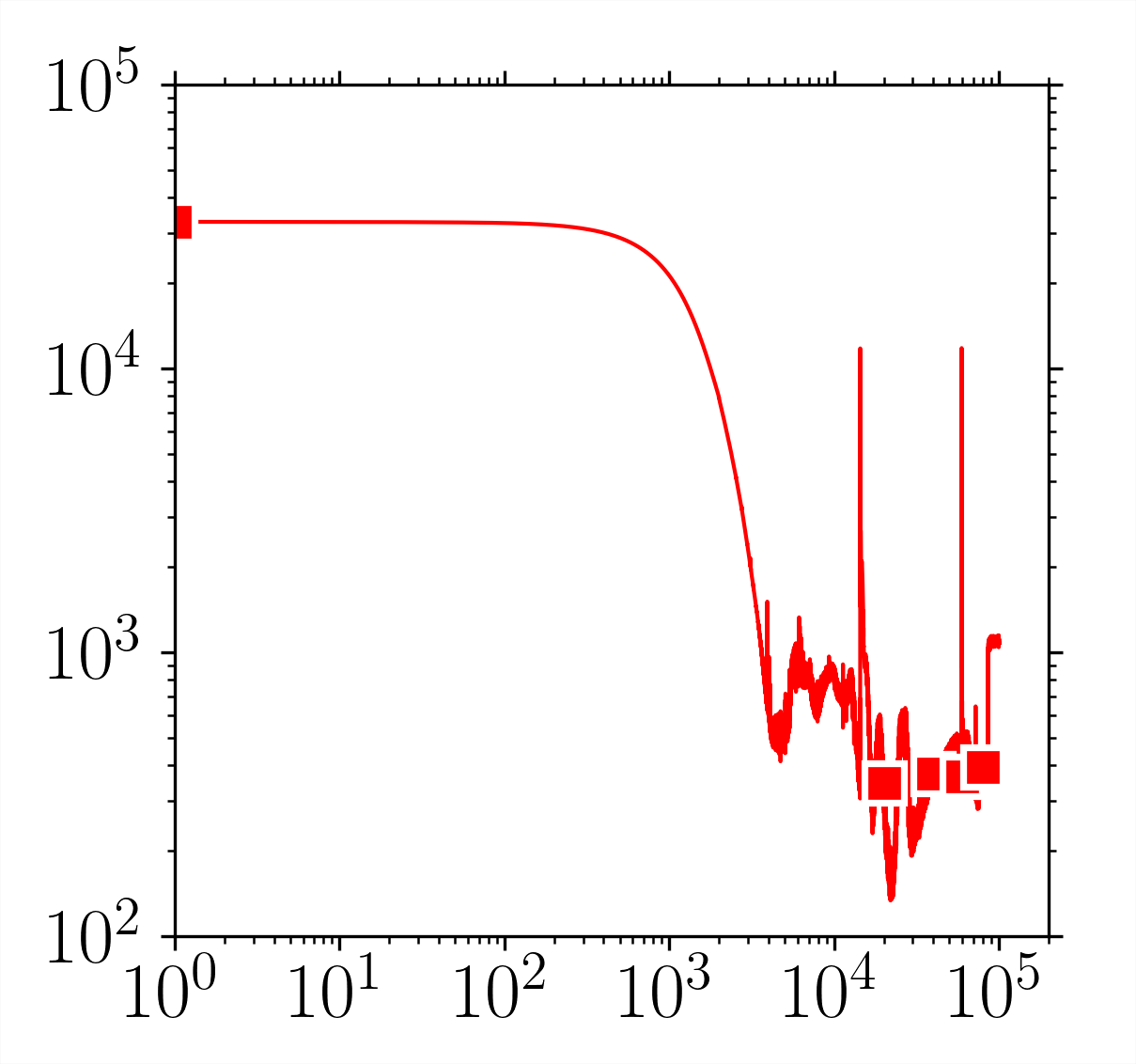}
        \LabelFig{17}{14}{$b)~\text{Test}$}
        \Labelxy{50}{-3}{0}{Epoch}
        \Labelxy{-3}{46}{90}{mse}
    \end{overpic}

    \begin{overpic}[width=0.23\textwidth]{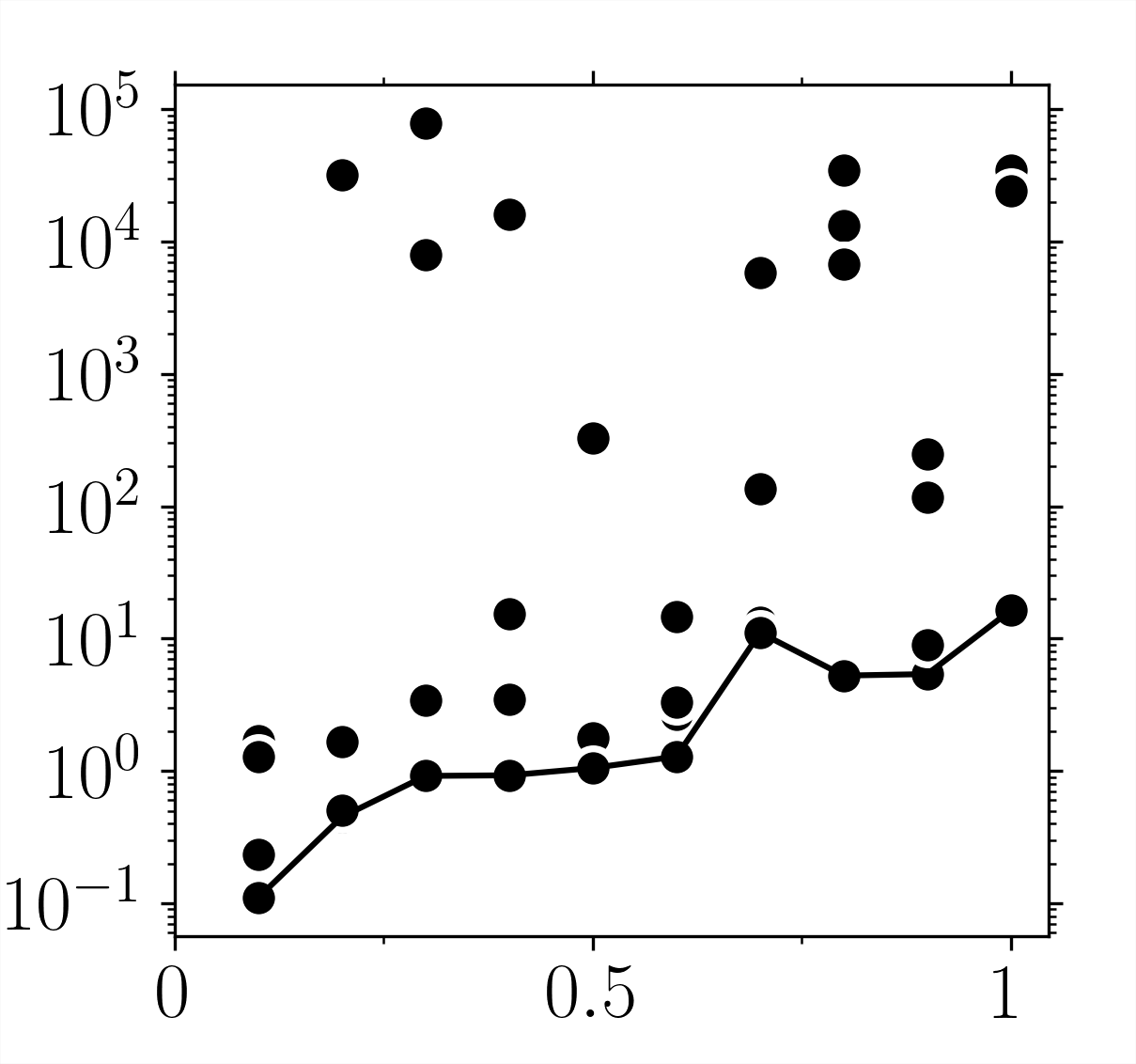}
        \LabelFig{17}{14}{$c)$}
        \Labelxy{50}{-3}{0}{Train. size}
        \Labelxy{-3}{46}{90}{mse}
    \end{overpic}
    \begin{overpic}[width=0.23\textwidth]{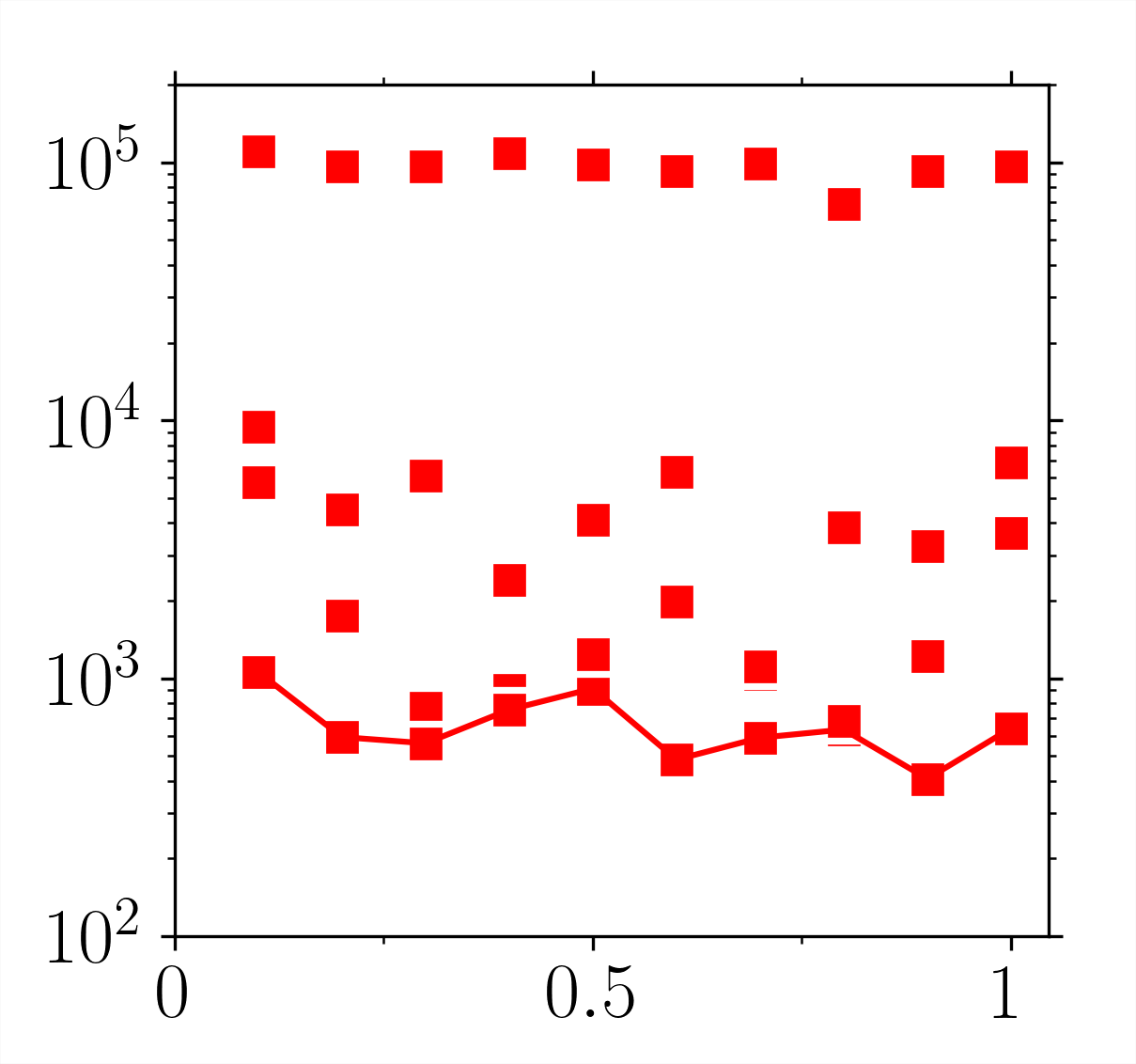}
        \LabelFig{17}{14}{$d)$}
        \Labelxy{50}{-3}{0}{Train. size}
        \Labelxy{-3}{46}{90}{mse}
    \end{overpic}
    \caption{Learning rate of the graph neural net corresponding to the  {a}) training  {b}) test set as a function of the number of iterations. Learning curves based on the cross-validated  {c}) training and  {d}) test mean squared errors for different training set sizes. The solid curves in  {c}) and  {d}) indicate lower bounds corresponding to the four-fold cross validation.}
    \label{fig:learning}
\end{figure}

The evolution of the GNN performance in minimizing the loss function is illustrated in Fig.~\ref{fig:learning}(a) and (b). The learning rate corresponding to the training and test data are shown as a function of the number of iterations.
The GNN training involves $92$ training cases ($70\%$) and $39$ test observations ($30\%$).
Applying the GNN model to the training data set lead to a fairly low training set error ($\text{MSE}<10$), showing a decay of at least four orders of magnitude after about $10^4$ iterations.
However, the test error rate appears to decrease more slowly as the optimization iterations proceed, decaying around two order of magnitude before it reaches a noise floor. 
Figure~\ref{fig:learning}(c) and (d) show learning curves for the GNN-based prediction task, plots of MSE against the (relative) size of the training set.
Here the results correspond to the four-fold cross-validation estimating the performance of the graph network over the training sets of varying sizes (and testing sets of a fixed size).
The performance tends to improves as the (relative) training set size increases to $50\%$ (less than 50 sample points) and increasing the number further leads to a small improvement.

\begin{figure}[b]
    \centering
    %
    \begin{overpic}[width=0.23\textwidth]{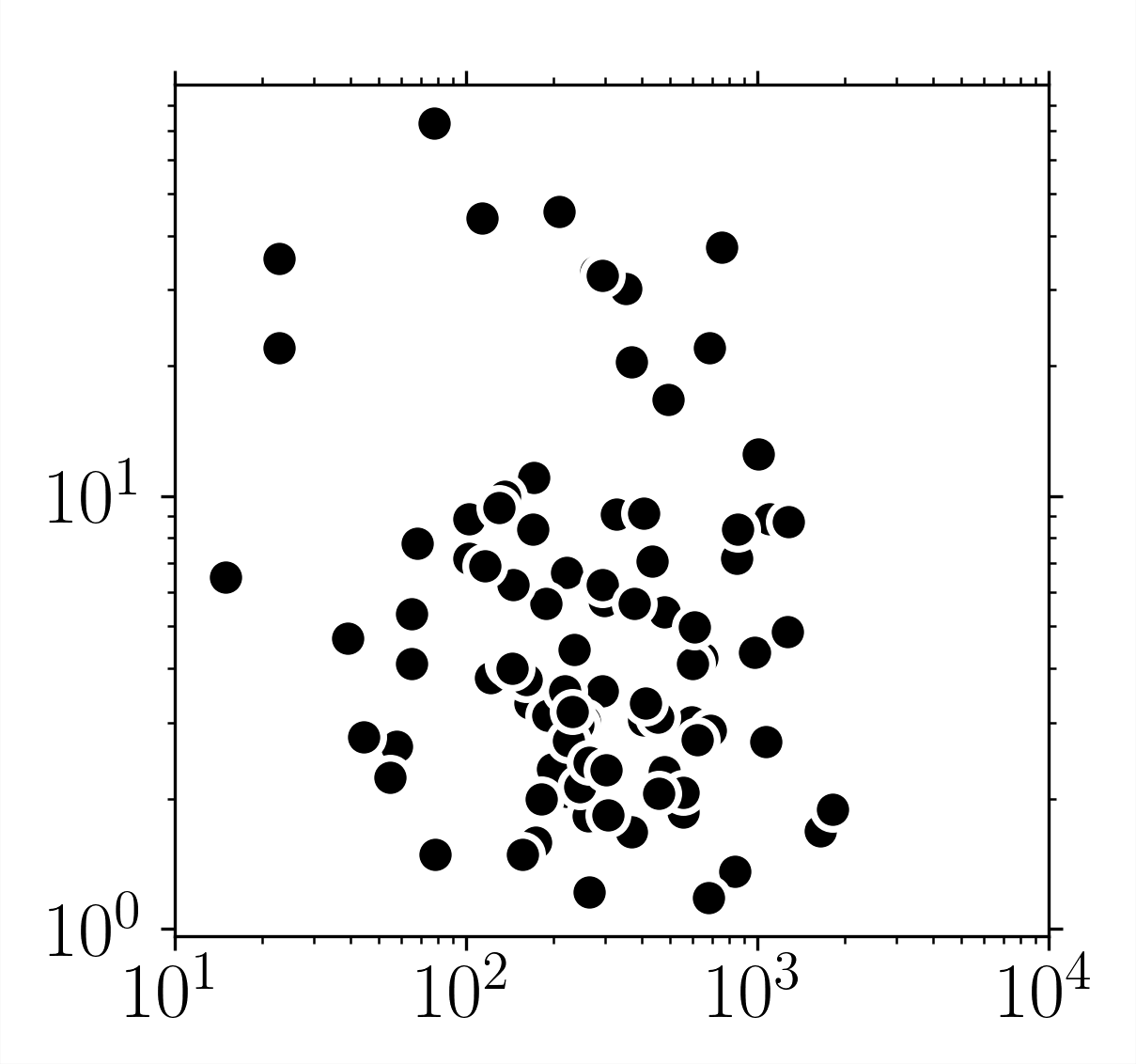}
        \Labelxy{50}{-3}{0}{\texttt{area}}
        \LabelFig{17}{14}{$a)~\text{Train}$}
        \Labelxy{-3}{46}{90}{mse}
    \end{overpic}
    \begin{overpic}[width=0.23\textwidth]{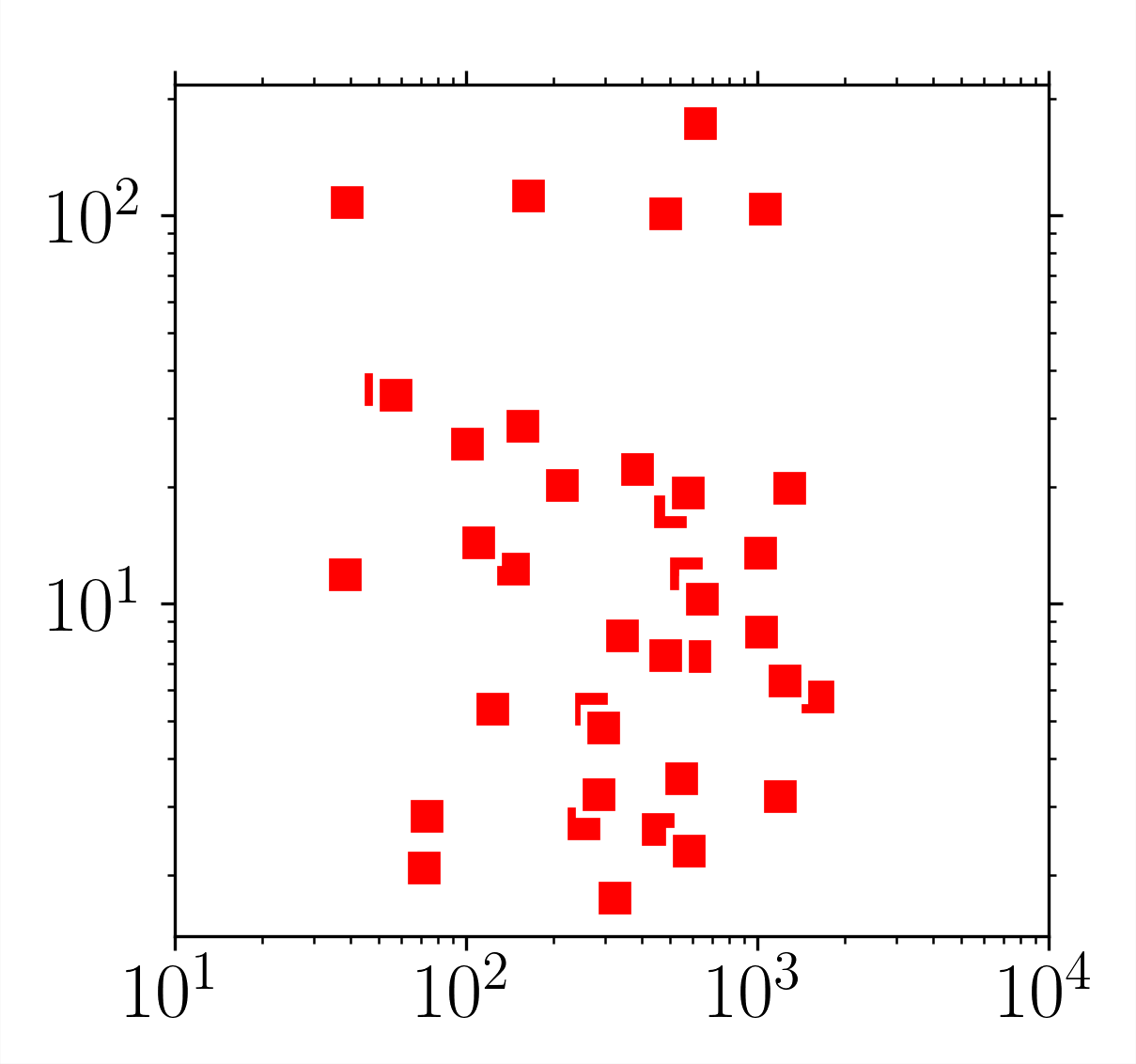}
        \Labelxy{50}{-3}{0}{\texttt{area}}
        \LabelFig{17}{14}{$b)~\text{Test}$}
        \Labelxy{-3}{46}{90}{mse}
    \end{overpic}
    %
    %
    \begin{overpic}[width=0.23\textwidth]{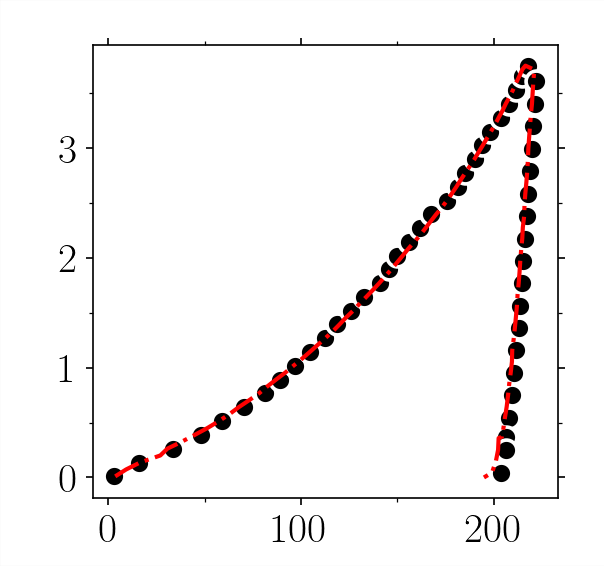}
        \put(14,44){\includegraphics[width=0.1\textwidth]{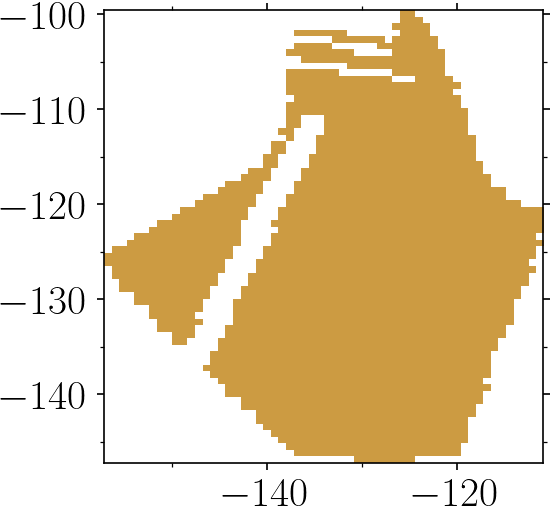}}
        \LabelFig{17}{14}{$c)~\scriptstyle\text{mse}\simeq 3,\texttt{area}\simeq 1200~\mu\text{m}^2$}
       \Labelxy{42}{-6}{0}{Depth(nm)}
        \Labelxy{-5}{32}{90}{{Load(mN)}}
     \end{overpic}
    %
    %
    \begin{overpic}[width=0.23\textwidth]{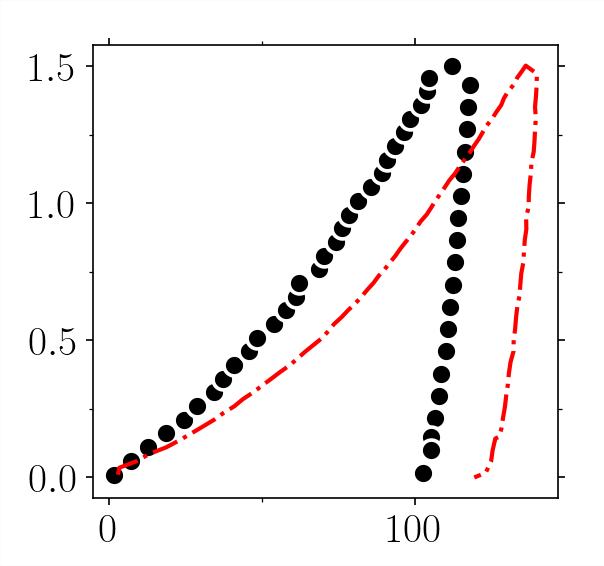}
        \put(14,66){\includegraphics[width=0.1\textwidth]{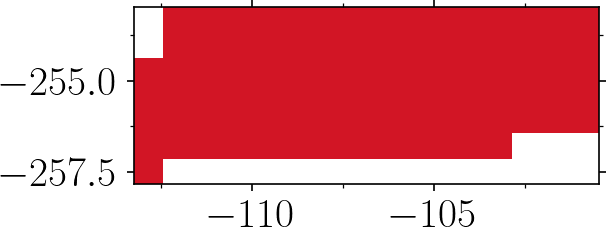}}
        \LabelFig{17}{14}{$d)~\scriptstyle\text{mse}\simeq 10^2,\texttt{area}\simeq 40~\mu\text{m}^2$}
       \Labelxy{42}{-6}{0}{Depth(nm)}
        \Labelxy{-5}{32}{90}{{Load(mN)}}
     \end{overpic}
    \caption{ {a}) Training and  {b}) test errors plotted against \texttt{area}. Actual (symbols) and predicted (dashdotted curve) load-depth curves associated with a  {c}) good and  {d}) poor prediction within the test set. The insets show the corresponding grain maps.}
    \label{fig:mse_grainSize}
\end{figure}

The predictive power of GNN as a function of variable \texttt{area} is shown in Fig.~\ref{fig:mse_grainSize} with the training and testing sets in Fig.~\ref{fig:mse_grainSize}(a) and Fig.~\ref{fig:mse_grainSize}(b)\remove{featuring statistically similar trends over a fairly broad range of scales}. \note[KK]{DONT GET THIS. revised.} 
As expected, the latter displays larger variations in terms of the test errors (almost three orders of magnitude in MSE).
The observed (anti-)correlations in both data sets indicate that, on average, the GNN exhibits a better performance with increasing grain size --- cf. Fig.~\ref{fig:mse_grainSize}(c), Fig.~\ref{fig:mse_grainSize}(d), Fig.~\ref{fig:prediction_test}, and Fig.~\ref{fig:prediction_train}.

\begin{figure}
    \centering
    %
    \begin{overpic}[width=0.23\textwidth]{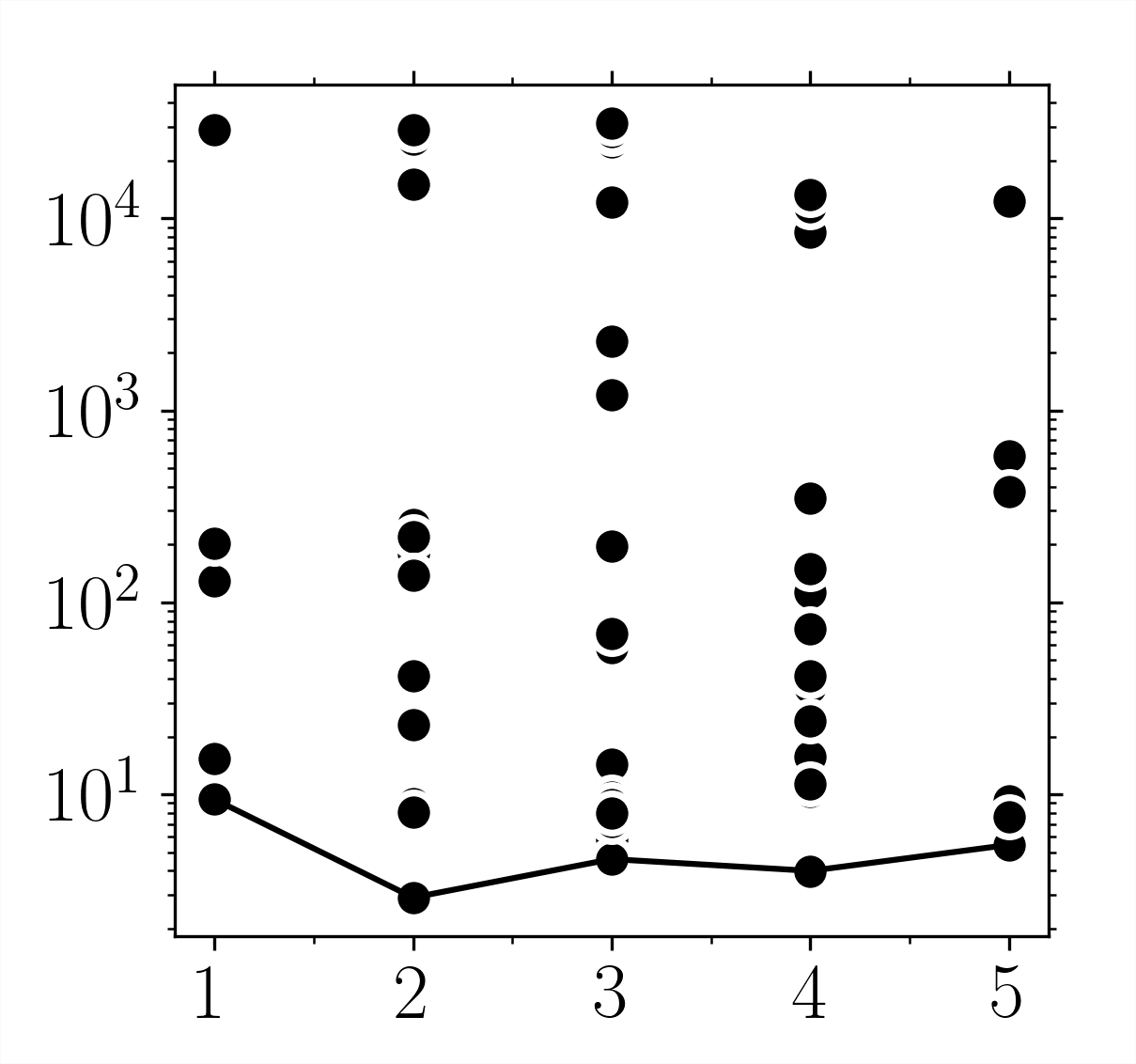}
        \Labelxy{50}{-3}{0}{subset size $k$}
        \LabelFig{17}{14}{$a)~\text{Train}$}
        \Labelxy{-3}{46}{90}{mse}
    \end{overpic}
    \begin{overpic}[width=0.23\textwidth]{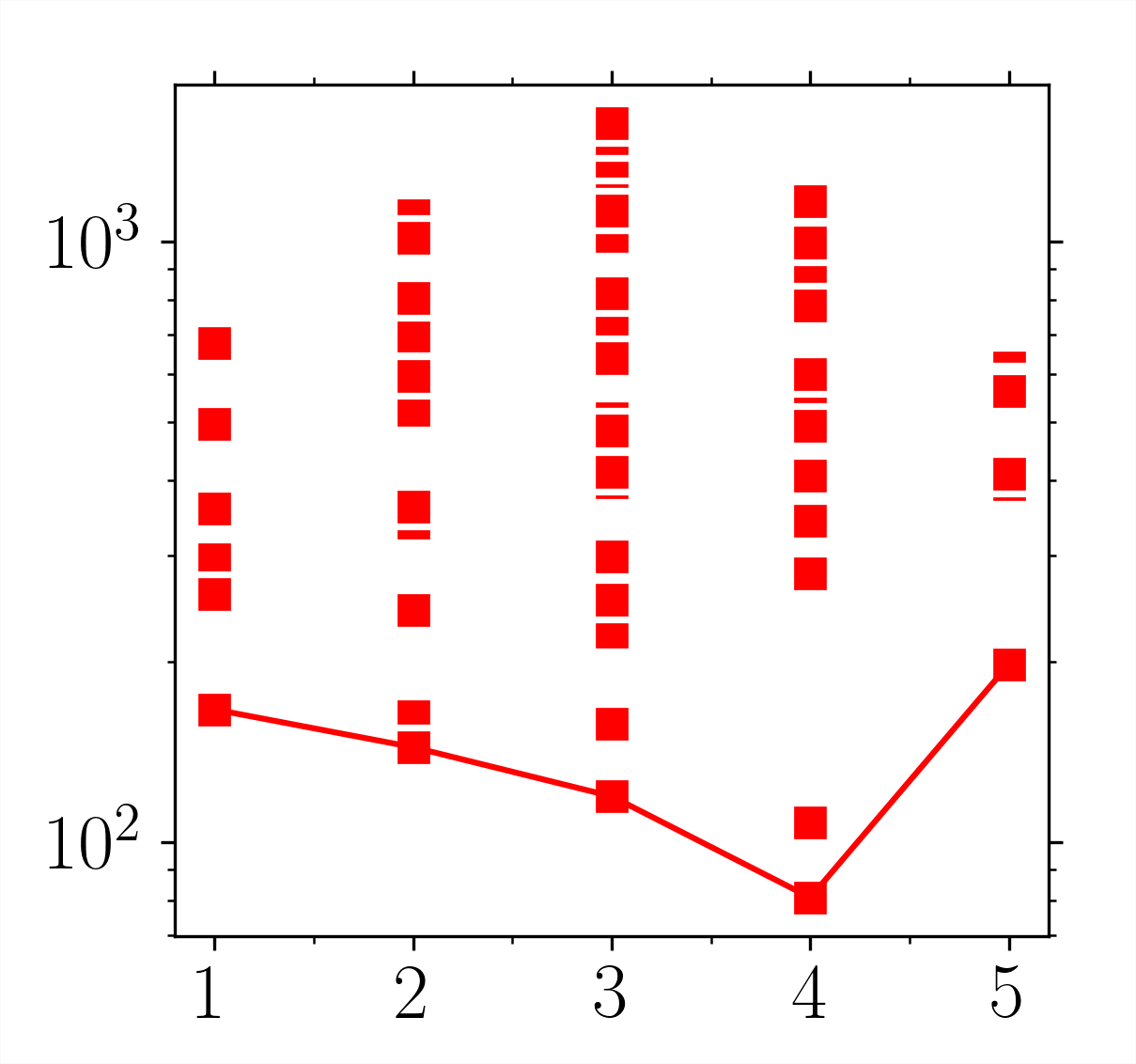}
        \Labelxy{50}{-3}{0}{subset size $k$}
        \LabelFig{17}{14}{$b)~\text{Test}$}
        \Labelxy{-3}{46}{90}{mse}
    \end{overpic}
    \vspace{12pt}

    \begin{minipage}{\columnwidth} 
            \centering
    \setlength\tabcolsep{6pt}
    \begin{tabularx}{\linewidth}{|p{3.2cm}|p{.5cm}|p{.5cm}|p{.5cm}|p{.5cm}|p{.5cm}} 
    \hline\hline
    \begin{tabular}{@{}l@{}} {subset} \\  {size} $k$  \end{tabular} &
    \begin{tabular}{@{}l@{}} $1$ \\  \end{tabular} &
    \begin{tabular}{@{}l@{}} $2$ \\  \end{tabular} &
    \begin{tabular}{@{}l@{}} $3$ \\  \end{tabular} &
    \begin{tabular}{@{}l@{}} $4$ \\  \end{tabular} &
    \begin{tabular}{@{}l@{}} $5$ \\  \end{tabular} \\
    \hline\hline  
    \texttt{area}   &  &  &$\dagger$ & &$\dagger$  \\
    \texttt{perimeter}   &  & &$\dagger$ &$\dagger$ &$\dagger$  \\
    \texttt{diameter} & $\dagger$ & &$\dagger$ &$\dagger$ &$\dagger$  \\
    \texttt{equivalentPerimeter}  &  &$\dagger$ & &$\dagger$ &$\dagger$ \\
    \texttt{shapeFactor} &  & &  & &$\dagger$  \\
    \texttt{numNeighbors} &  &$\dagger$ & &$\dagger$ & \\
\hline
        \end{tabularx}
        \end{minipage}
\caption{Validation curves determining  {a}) training and  {b}) test errors for varying sets of descriptors. Here symbols correspond to all possible subsets of size $k=1...5$ corresponding to the full set of numeric variables \texttt{area}, \texttt{perimeter}, \texttt{diameter}, \texttt{equivalentPerimeter}, \texttt{shapeFactor}, \texttt{numNeighbors}. The solid curves indicate minimal errors corresponding to each $k$ and the table denotes the associated set of predictors (relevant to the testing set) by $\dagger$.}
    \label{fig:validation_curve}
\end{figure}

 Figure~\ref{fig:validation_curve} features the performance of the GNN model for all the predictor subsets (excluding the binary metrics).
We pre-selected six quantitative variables including \texttt{area}, \texttt{perimeter}, \texttt{diameter}, \texttt{equivalentPerimeter}, \texttt{shapeFactor}, and \texttt{numNeighbors} to probe the training and test errors for every possible subset of size $k=1...5$.
In every analysis, the nodal coordinates (\texttt{x}, \texttt{y}) as well as the edge-based predictors \texttt{misOrientationAngle} and \texttt{boundaryLength} are incorporated as a fixed set of descriptors that are supplemented by additional variables as described above.
Overall, the training errors in Fig.~\ref{fig:validation_curve}(a) and the associated base-line do not seem to be very sensitive to the subset size $k$. 
However, the minimum MSE corresponding to the testing set in Fig.~\ref{fig:validation_curve}(b) shows meaningful variations with $k$ featuring a dip at $k=4$ that corresponds to \texttt{perimeter}, \texttt{diameter}, \texttt{equivalentPerimeter}, and \texttt{numNeighbors} as predictor variables (see the table). 
\note[KK]{WOULD BE GOOD TO TRANSLATE THIS INTO ACTUAL QUANTITTES NOT NAMES OF DESCRIPTORS.
Using  the first letters or acronyms for this large set of variables would not be very descriptive!}
In fact, the model performance will drastically degrade by including a subset of size $k=5$.
Out of the six numeric variables, \texttt{diameter} and \texttt{shapeFactor} are the most and least repeated entries of the table in Fig.~\ref{fig:validation_curve} and, therefore, can be viewed as the most and least relevant descriptors.

\subsection{Hardness Prediction}
\begin{figure}[b]
    \begin{overpic}[width=0.24\textwidth]{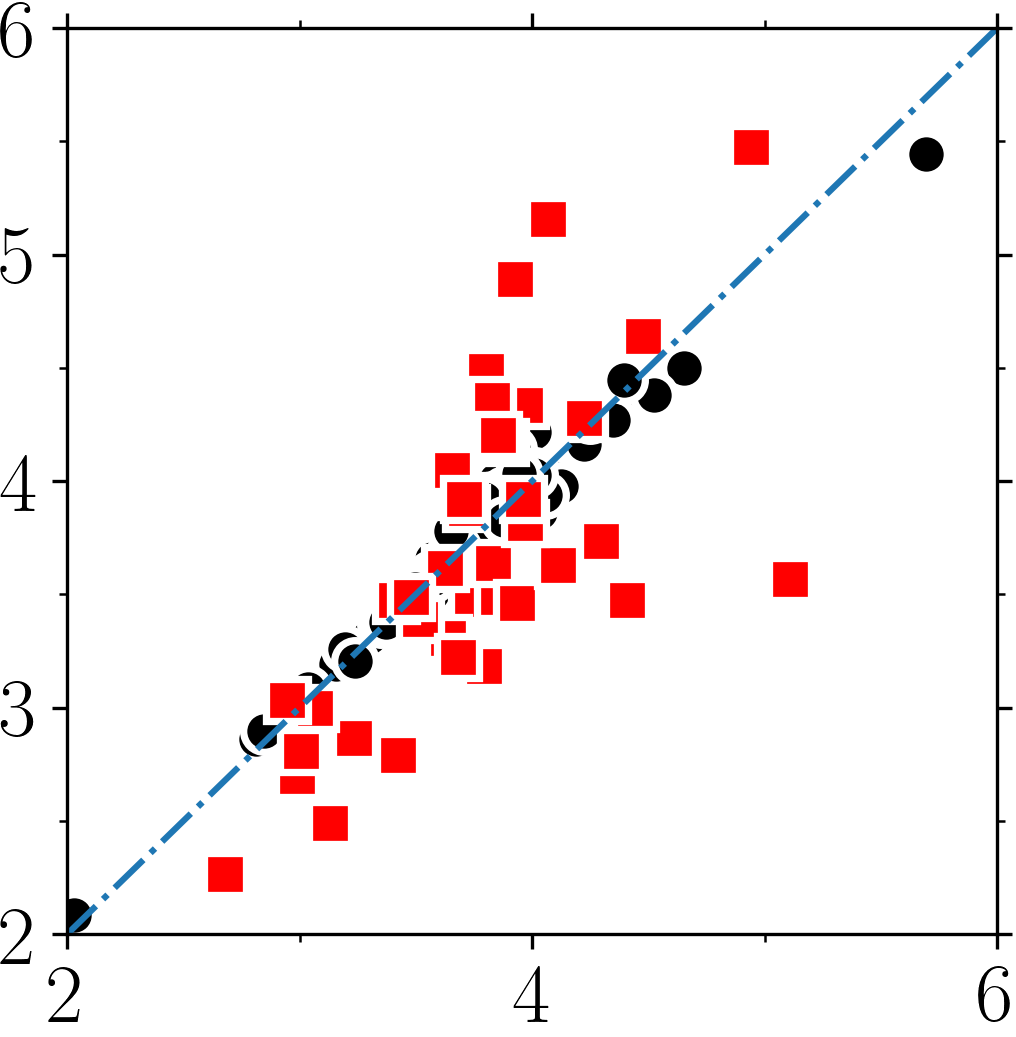}
        \Labelxy{50}{-6}{0}{$h_\text{act}$(Gpa)}
        \Labelxy{-12}{50}{90}{$h_\text{pred}$(Gpa)}
    \end{overpic}
    \caption{Scatter plot of the predicted hardness and the actual values corresponding to the training set (\protect\circTxtFill{1}{1}{black}) and the test set (\protect\legSqTxt{1}{1}{red}). The diagonal dashdotted line indicates $h_\text{pred}=h_\text{act}$. The hardness is measured in Gpa.}
    \label{fig:h_scatter}
\end{figure}

\begin{figure*}[t]
    \begin{overpic}[width=0.28\textwidth]{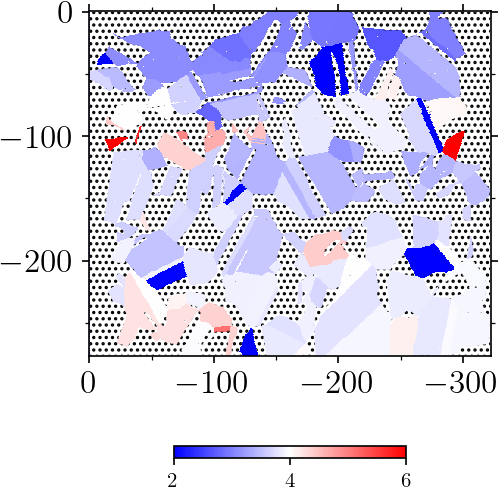}
        \LabelFig{16}{99}{$a)~h_\text{act}$}
        \Labelxy{50}{13}{0}{$x(\mu \text{m})$}
        \Labelxy{-8}{50}{90}{$y(\mu \text{m})$}
    \end{overpic}
    \begin{overpic}[width=0.28\textwidth]{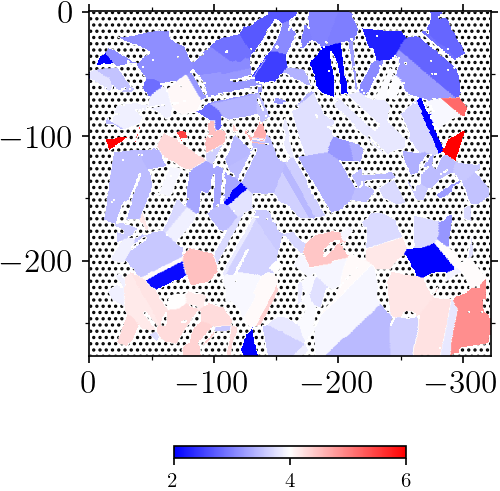}
        \LabelFig{16}{99}{$b)~ h_\text{pred}$}
        \Labelxy{50}{13}{0}{$x(\mu \text{m})$}
    \end{overpic}
    \begin{overpic}[width=0.28\textwidth]{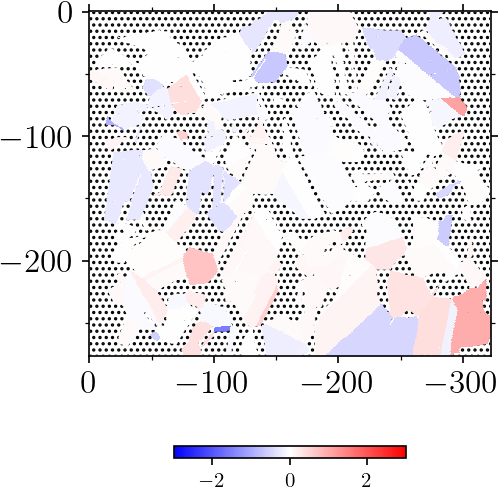}
        \LabelFig{16}{99}{$c)~ h_\text{pred}-h_\text{act}$}
        \Labelxy{50}{13}{0}{$x(\mu \text{m})$}
    \end{overpic}
    \caption{Hardness maps (including indented grains only) associated with the  {a}) actual data $h_\text{act}$  {b}) GNN prediction $h_\text{pred}$  {c}) difference between predicted and actual data $h_\text{pred}-H_\text{act}$. The harness is measured in Gpa. The hatched areas denote non-indented grains.}
    \label{fig:hmap}
\end{figure*}

We infer grain-scale hardness from the GNN-predicted load-depth diagrams following the Oliver-Pharr framework \cite{oliver1992improved}.
The scatter plot of the predicted ($h_\text{pred}$) and actual hardness ($h_\text{act}$) is also shown in Fig.~\ref{fig:h_scatter}.
The predictions associated with the testing set is, apart from a few outliers, reasonably distributed around $h_\text{pred}=h_\text{act}$. 
This is further quantified by a fairly high Pearson's correlation coefficient $\rho_h=\langle \hat{h}_\text{pred}~\hat{h}_\text{act}\rangle$  with $\rho^\text{test}_h\simeq 0.7$.
Here $\hat{h}\doteq(h-\langle h \rangle)/\text{var}^\frac{1}{2}(h)$ with the angular brackets $\langle . \rangle$ denoting an average.
The hardness maps associated with the indented grains are shown for the actual and predicted data sets in in Fig.~\ref{fig:hmap}(a) and (b) as well as the difference between the former and the latter as in in Fig.~\ref{fig:hmap}(c).
The actual map in Fig.~\ref{fig:hmap}(a) indicates that harder grains (in red) are, on average, smaller in size.  
We note that the bluish (reddish) colors in Fig.~\ref{fig:hmap}(c) indicate regions where the GNN tends to over(under)predict hardness.  
\add[KK]{A visual inspection of Fig.~\ref{fig:hmap}(c) might indicate that grains with over(under)-predicted hardness may not possess any clear, distinguishable features (i.e. an abnormal geometry) that could delineate these cases.
In this context, dimensionality reduction and/or clustering algorithms might allow for extracting anomalous attributes associated with such outliers (cf. Fig.~\ref{fig:h_scatter}) in a more systematic way.}  
\note[KK]{OK, COMPARING THE FIGS 7 AND 8 POSE THE QUESTION, CAN ONE SAY SOMETHING ABOUT THE OUTLIERS (IN PARTICULAR FOR HPRED OVER 4 GPA), ARE THERE ANY FEATURES THAT DISTINGUISH THOSE CASES WHICH OVER/UNDERSHOOT?
very good point! this is the whole idea behind fig. 8(c). indeed, there *might* be characteristic features that could explain this over/undershoot. however, such features are not visually apparent. To probe it in a more systematic way, one could look at a high-dimensional feature space and seek for possible clustering based on the amount of overshoot (like what Henri did in his paper but in a supervised manner). it's very likely that referees would ask for such an analysis.}

\section{Conclusions \& Discussions }
We have studied the effectiveness and robustness of a GNN-based supervised machine learning model in predicting mechanical nanoindentation response from experimentally-measured grain microstructure replicated as a graph.
Microstructural patterns are encapsulated in the GNN via a set of node-based and edge-based hidden layers that learn from nanoindentation-induced deformation in a supervised learning context.
We have probed hardness as an experimentally measurable micromechanical property to test the predictive power of the GNN.
A rich set of grain-level structural features was extracted from the grain map and the robustness and accuracy of the prediction task was verified with respect to varying subsets of selected descriptors. 

Given the convenience of GNNs (i.e. predictiveness, speed, and interpretability) in the hardness prediction, the proposed framework may also be augmented to account for indentation-induced pop-in behavior abrupt displacement jumps (in a load-controlled indentation) and associated statistical distributions solely based on microstructural inputs.
As pop-ins typically trigger as a result of the interplay between dislocations and embedding grain boundary, one might envision the use of more elaborate indicators of microstructure (such as dislocation density) to be incorporated as nodal and/or edge-based ingredients.

As a final remark, applications of data-driven methodologies shall not be regarded as substitutes but rather complements to laboratory-based measurements and/or high-throughput physics-based simulations. 
Machine-learned models require smooth access to well-maintained, accurate, and reusable data sets, relevant to materials' micro-structure and associated (micro-)mechanical response, which are otherwise impossible to measure in the absence of experimental/numeric observations. 
In fact, a coherent integration of the above methodologies will be essential in a way that they guide one another to achieve the desired speed, interpretability, and predictiveness of outcomes. 
Our GNN development provides a fine example in this context, where a fairly limited number of surface measurements (order $10^2$ indentation tests) was performed for the prediction task.
Nevertheless, the model outcomes will allow us to efficiently infer a \emph{full} hardness cartography map from \add[KK]{the prescribed force dynamics, as in Fig.~\ref{fig:force_time}(a)}, and a fine-scale EBSD analysis of nearly $10^3$ grains.
\note[KK]{UMMM. THIS HAS ACTUALLY NOT BEEN DONE... FIGURE 8 HAS ONLY THE TESTED GRAINS?? figure added in SM. yes. on top of the grain map, the force timeseries (as in Fig.~\ref{fig:force_time}(a)) is also needed to make hardness predictions. I have changed the above sentence accordingly.}

\begin{acknowledgments}
This research was funded by the European Union Horizon 2020 research and innovation program under grant agreement no. 857470 and from the European Regional Development Fund via Foundation for Polish Science International Research Agenda PLUS program grant no. MAB PLUS/2018/8. We wish to acknowledge fruitful discussions with Daniel Cieslinski.
\end{acknowledgments}

\bibliography{references}

\begin{thebibliography}{33}%
\makeatletter
\providecommand \@ifxundefined [1]{%
 \@ifx{#1\undefined}
}%
\providecommand \@ifnum [1]{%
 \ifnum #1\expandafter \@firstoftwo
 \else \expandafter \@secondoftwo
 \fi
}%
\providecommand \@ifx [1]{%
 \ifx #1\expandafter \@firstoftwo
 \else \expandafter \@secondoftwo
 \fi
}%
\providecommand \natexlab [1]{#1}%
\providecommand \enquote  [1]{``#1''}%
\providecommand \bibnamefont  [1]{#1}%
\providecommand \bibfnamefont [1]{#1}%
\providecommand \citenamefont [1]{#1}%
\providecommand \href@noop [0]{\@secondoftwo}%
\providecommand \href [0]{\begingroup \@sanitize@url \@href}%
\providecommand \@href[1]{\@@startlink{#1}\@@href}%
\providecommand \@@href[1]{\endgroup#1\@@endlink}%
\providecommand \@sanitize@url [0]{\catcode `\\12\catcode `\$12\catcode
  `\&12\catcode `\#12\catcode `\^12\catcode `\_12\catcode `\%12\relax}%
\providecommand \@@startlink[1]{}%
\providecommand \@@endlink[0]{}%
\providecommand \url  [0]{\begingroup\@sanitize@url \@url }%
\providecommand \@url [1]{\endgroup\@href {#1}{\urlprefix }}%
\providecommand \urlprefix  [0]{URL }%
\providecommand \Eprint [0]{\href }%
\providecommand \doibase [0]{https://doi.org/}%
\providecommand \selectlanguage [0]{\@gobble}%
\providecommand \bibinfo  [0]{\@secondoftwo}%
\providecommand \bibfield  [0]{\@secondoftwo}%
\providecommand \translation [1]{[#1]}%
\providecommand \BibitemOpen [0]{}%
\providecommand \bibitemStop [0]{}%
\providecommand \bibitemNoStop [0]{.\EOS\space}%
\providecommand \EOS [0]{\spacefactor3000\relax}%
\providecommand \BibitemShut  [1]{\csname bibitem#1\endcsname}%
\let\auto@bib@innerbib\@empty
\bibitem [{\citenamefont {Wadhwa}\ and\ \citenamefont
  {Dhaliwal}(2008)}]{wadhwa2008textbook}%
  \BibitemOpen
  \bibfield  {author} {\bibinfo {author} {\bibfnamefont {A.~S.}\ \bibnamefont
  {Wadhwa}}\ and\ \bibinfo {author} {\bibfnamefont {H.~S.}\ \bibnamefont
  {Dhaliwal}},\ }\href@noop {} {\emph {\bibinfo {title} {A Textbook of
  Engineering Material and Metallurgy}}}\ (\bibinfo  {publisher} {Firewall
  Media},\ \bibinfo {year} {2008})\BibitemShut {NoStop}%
\bibitem [{\citenamefont {Lasalmonie}\ and\ \citenamefont
  {Strudel}(1986)}]{lasalmonie1986influence}%
  \BibitemOpen
  \bibfield  {author} {\bibinfo {author} {\bibfnamefont {A.}~\bibnamefont
  {Lasalmonie}}\ and\ \bibinfo {author} {\bibfnamefont {J.}~\bibnamefont
  {Strudel}},\ }\bibfield  {title} {\bibinfo {title} {Influence of grain size
  on the mechanical behaviour of some high strength materials},\ }\href@noop {}
  {\bibfield  {journal} {\bibinfo  {journal} {Journal of Materials Science}\
  }\textbf {\bibinfo {volume} {21}},\ \bibinfo {pages} {1837} (\bibinfo {year}
  {1986})}\BibitemShut {NoStop}%
\bibitem [{\citenamefont {Hall}(1951)}]{hall1951deformation}%
  \BibitemOpen
  \bibfield  {author} {\bibinfo {author} {\bibfnamefont {E.}~\bibnamefont
  {Hall}},\ }\bibfield  {title} {\bibinfo {title} {The deformation and ageing
  of mild steel: Iii discussion of results},\ }\href@noop {} {\bibfield
  {journal} {\bibinfo  {journal} {Proceedings of the Physical Society. Section
  B}\ }\textbf {\bibinfo {volume} {64}},\ \bibinfo {pages} {747} (\bibinfo
  {year} {1951})}\BibitemShut {NoStop}%
\bibitem [{\citenamefont {Petch}(1953)}]{petch1953cleavage}%
  \BibitemOpen
  \bibfield  {author} {\bibinfo {author} {\bibfnamefont {N.}~\bibnamefont
  {Petch}},\ }\bibfield  {title} {\bibinfo {title} {The cleavage strength of
  polycrystals},\ }\href@noop {} {\bibfield  {journal} {\bibinfo  {journal}
  {Journal of the Iron and Steel institute}\ }\textbf {\bibinfo {volume}
  {174}},\ \bibinfo {pages} {25} (\bibinfo {year} {1953})}\BibitemShut
  {NoStop}%
\bibitem [{\citenamefont {Alavudeen}\ \emph {et~al.}(2006)\citenamefont
  {Alavudeen}, \citenamefont {Venkateshwaran},\ and\ \citenamefont
  {Jappes}}]{alavudeen2006textbook}%
  \BibitemOpen
  \bibfield  {author} {\bibinfo {author} {\bibfnamefont {A.}~\bibnamefont
  {Alavudeen}}, \bibinfo {author} {\bibfnamefont {N.}~\bibnamefont
  {Venkateshwaran}},\ and\ \bibinfo {author} {\bibfnamefont {J.~W.}\
  \bibnamefont {Jappes}},\ }\href@noop {} {\emph {\bibinfo {title} {A textbook
  of engineering materials and metallurgy}}}\ (\bibinfo  {publisher} {Firewall
  Media},\ \bibinfo {year} {2006})\BibitemShut {NoStop}%
\bibitem [{\citenamefont {Yang}\ and\ \citenamefont
  {Vehoff}(2005)}]{yang2005grain}%
  \BibitemOpen
  \bibfield  {author} {\bibinfo {author} {\bibfnamefont {B.}~\bibnamefont
  {Yang}}\ and\ \bibinfo {author} {\bibfnamefont {H.}~\bibnamefont {Vehoff}},\
  }\bibfield  {title} {\bibinfo {title} {Grain size effects on the mechanical
  properties of nanonickel examined by nanoindentation},\ }\href@noop {}
  {\bibfield  {journal} {\bibinfo  {journal} {Materials Science and
  Engineering: A}\ }\textbf {\bibinfo {volume} {400}},\ \bibinfo {pages} {467}
  (\bibinfo {year} {2005})}\BibitemShut {NoStop}%
\bibitem [{\citenamefont {Wang}\ and\ \citenamefont
  {Ngan}(2004)}]{wang2004indentation}%
  \BibitemOpen
  \bibfield  {author} {\bibinfo {author} {\bibfnamefont {M.}~\bibnamefont
  {Wang}}\ and\ \bibinfo {author} {\bibfnamefont {A.}~\bibnamefont {Ngan}},\
  }\bibfield  {title} {\bibinfo {title} {Indentation strain burst phenomenon
  induced by grain boundaries in niobium},\ }\href@noop {} {\bibfield
  {journal} {\bibinfo  {journal} {Journal of materials research}\ }\textbf
  {\bibinfo {volume} {19}},\ \bibinfo {pages} {2478} (\bibinfo {year}
  {2004})}\BibitemShut {NoStop}%
\bibitem [{\citenamefont {Britton}\ \emph {et~al.}(2009)\citenamefont
  {Britton}, \citenamefont {Randman},\ and\ \citenamefont
  {Wilkinson}}]{britton2009nanoindentation}%
  \BibitemOpen
  \bibfield  {author} {\bibinfo {author} {\bibfnamefont {T.}~\bibnamefont
  {Britton}}, \bibinfo {author} {\bibfnamefont {D.}~\bibnamefont {Randman}},\
  and\ \bibinfo {author} {\bibfnamefont {A.}~\bibnamefont {Wilkinson}},\
  }\bibfield  {title} {\bibinfo {title} {Nanoindentation study of slip transfer
  phenomenon at grain boundaries},\ }\href@noop {} {\bibfield  {journal}
  {\bibinfo  {journal} {Journal of Materials Research}\ }\textbf {\bibinfo
  {volume} {24}},\ \bibinfo {pages} {607} (\bibinfo {year} {2009})}\BibitemShut
  {NoStop}%
\bibitem [{\citenamefont {Pathak}\ \emph {et~al.}(2012)\citenamefont {Pathak},
  \citenamefont {Michler}, \citenamefont {Wasmer},\ and\ \citenamefont
  {Kalidindi}}]{pathak2012studying}%
  \BibitemOpen
  \bibfield  {author} {\bibinfo {author} {\bibfnamefont {S.}~\bibnamefont
  {Pathak}}, \bibinfo {author} {\bibfnamefont {J.}~\bibnamefont {Michler}},
  \bibinfo {author} {\bibfnamefont {K.}~\bibnamefont {Wasmer}},\ and\ \bibinfo
  {author} {\bibfnamefont {S.~R.}\ \bibnamefont {Kalidindi}},\ }\bibfield
  {title} {\bibinfo {title} {Studying grain boundary regions in polycrystalline
  materials using spherical nano-indentation and orientation imaging
  microscopy},\ }\href@noop {} {\bibfield  {journal} {\bibinfo  {journal}
  {Journal of Materials Science}\ }\textbf {\bibinfo {volume} {47}},\ \bibinfo
  {pages} {815} (\bibinfo {year} {2012})}\BibitemShut {NoStop}%
\bibitem [{\citenamefont {Ohmura}\ \emph {et~al.}(2005)\citenamefont {Ohmura},
  \citenamefont {Tsuzaki},\ and\ \citenamefont
  {Yin}}]{ohmura2005nanoindentation}%
  \BibitemOpen
  \bibfield  {author} {\bibinfo {author} {\bibfnamefont {T.}~\bibnamefont
  {Ohmura}}, \bibinfo {author} {\bibfnamefont {K.}~\bibnamefont {Tsuzaki}},\
  and\ \bibinfo {author} {\bibfnamefont {F.}~\bibnamefont {Yin}},\ }\bibfield
  {title} {\bibinfo {title} {Nanoindentation-induced deformation behavior in
  the vicinity of single grain boundary of interstitial-free steel},\
  }\href@noop {} {\bibfield  {journal} {\bibinfo  {journal} {Materials
  transactions}\ }\textbf {\bibinfo {volume} {46}},\ \bibinfo {pages} {2026}
  (\bibinfo {year} {2005})}\BibitemShut {NoStop}%
\bibitem [{\citenamefont {Ozturk}\ \emph {et~al.}(2015)\citenamefont {Ozturk},
  \citenamefont {Stein}, \citenamefont {Pokharel}, \citenamefont {Hefferan},
  \citenamefont {Tucker}, \citenamefont {Jha}, \citenamefont {John},
  \citenamefont {Lebensohn}, \citenamefont {Kenesei}, \citenamefont {Suter},\
  and\ \citenamefont {Rollett}}]{grain-1}%
  \BibitemOpen
  \bibfield  {author} {\bibinfo {author} {\bibfnamefont {T.}~\bibnamefont
  {Ozturk}}, \bibinfo {author} {\bibfnamefont {C.}~\bibnamefont {Stein}},
  \bibinfo {author} {\bibfnamefont {R.}~\bibnamefont {Pokharel}}, \bibinfo
  {author} {\bibfnamefont {C.}~\bibnamefont {Hefferan}}, \bibinfo {author}
  {\bibfnamefont {H.}~\bibnamefont {Tucker}}, \bibinfo {author} {\bibfnamefont
  {S.~K.}\ \bibnamefont {Jha}}, \bibinfo {author} {\bibfnamefont
  {R.}~\bibnamefont {John}}, \bibinfo {author} {\bibfnamefont {R.~A.}\
  \bibnamefont {Lebensohn}}, \bibinfo {author} {\bibfnamefont {P.}~\bibnamefont
  {Kenesei}}, \bibinfo {author} {\bibfnamefont {R.~M.}\ \bibnamefont {Suter}},\
  and\ \bibinfo {author} {\bibfnamefont {A.~D.}\ \bibnamefont {Rollett}},\
  }\bibfield  {title} {\bibinfo {title} {Simulation domain size requirements
  for elastic response of 3d polycrystalline materials},\ }\href@noop {}
  {\bibfield  {journal} {\bibinfo  {journal} {Modelling and Simulation in
  Materials Science and Engineering}\ }\textbf {\bibinfo {volume} {24}}
  (\bibinfo {year} {2015})}\BibitemShut {NoStop}%
\bibitem [{\citenamefont {Jackson}\ \emph {et~al.}(2019)\citenamefont
  {Jackson}, \citenamefont {Groeber}, \citenamefont {Donegan},\ and\
  \citenamefont {Dimiduk}}]{grain-2}%
  \BibitemOpen
  \bibfield  {author} {\bibinfo {author} {\bibfnamefont {M.}~\bibnamefont
  {Jackson}}, \bibinfo {author} {\bibfnamefont {M.}~\bibnamefont {Groeber}},
  \bibinfo {author} {\bibfnamefont {S.}~\bibnamefont {Donegan}},\ and\ \bibinfo
  {author} {\bibfnamefont {D.}~\bibnamefont {Dimiduk}},\ }\bibfield  {title}
  {\bibinfo {title} {Advancements to the digital representation environment for
  analysis of materials in 3-dimensions—dream. 3d},\ }\href@noop {}
  {\bibfield  {journal} {\bibinfo  {journal} {Microscopy and Microanalysis}\
  }\textbf {\bibinfo {volume} {25}},\ \bibinfo {pages} {176} (\bibinfo {year}
  {2019})}\BibitemShut {NoStop}%
\bibitem [{\citenamefont {Bronkhorst}\ \emph {et~al.}(1992)\citenamefont
  {Bronkhorst}, \citenamefont {Kalidindi},\ and\ \citenamefont
  {Anand}}]{grain-3}%
  \BibitemOpen
  \bibfield  {author} {\bibinfo {author} {\bibfnamefont {C.~A.}\ \bibnamefont
  {Bronkhorst}}, \bibinfo {author} {\bibfnamefont {S.}~\bibnamefont
  {Kalidindi}},\ and\ \bibinfo {author} {\bibfnamefont {L.}~\bibnamefont
  {Anand}},\ }\bibfield  {title} {\bibinfo {title} {Polycrystalline plasticity
  and the evolution of crystallographic texture in fcc metals},\ }\href@noop {}
  {\bibfield  {journal} {\bibinfo  {journal} {Philosophical Transactions of the
  Royal Society of London. Series A: Physical and Engineering Sciences}\
  }\textbf {\bibinfo {volume} {341}},\ \bibinfo {pages} {443} (\bibinfo {year}
  {1992})}\BibitemShut {NoStop}%
\bibitem [{\citenamefont {Bronkhorst}\ \emph {et~al.}(2019)\citenamefont
  {Bronkhorst}, \citenamefont {Mayeur}, \citenamefont {Livescu}, \citenamefont
  {Pokharel}, \citenamefont {Brown},\ and\ \citenamefont {Gray~III}}]{grain-4}%
  \BibitemOpen
  \bibfield  {author} {\bibinfo {author} {\bibfnamefont {C.~A.}\ \bibnamefont
  {Bronkhorst}}, \bibinfo {author} {\bibfnamefont {J.~R.}\ \bibnamefont
  {Mayeur}}, \bibinfo {author} {\bibfnamefont {V.}~\bibnamefont {Livescu}},
  \bibinfo {author} {\bibfnamefont {R.}~\bibnamefont {Pokharel}}, \bibinfo
  {author} {\bibfnamefont {D.~W.}\ \bibnamefont {Brown}},\ and\ \bibinfo
  {author} {\bibfnamefont {G.~T.}\ \bibnamefont {Gray~III}},\ }\bibfield
  {title} {\bibinfo {title} {Structural representation of additively
  manufactured 316l austenitic stainless steel},\ }\href@noop {} {\bibfield
  {journal} {\bibinfo  {journal} {International Journal of Plasticity}\
  }\textbf {\bibinfo {volume} {118}},\ \bibinfo {pages} {70} (\bibinfo {year}
  {2019})}\BibitemShut {NoStop}%
\bibitem [{\citenamefont {Battaglia}\ \emph {et~al.}(2018)\citenamefont
  {Battaglia}, \citenamefont {Hamrick}, \citenamefont {Bapst}, \citenamefont
  {Sanchez-Gonzalez}, \citenamefont {Zambaldi}, \citenamefont {Malinowski},
  \citenamefont {Tacchetti}, \citenamefont {Raposo}, \citenamefont {Santoro},
  \citenamefont {Faulkner} \emph {et~al.}}]{battaglia2018relational}%
  \BibitemOpen
  \bibfield  {author} {\bibinfo {author} {\bibfnamefont {P.~W.}\ \bibnamefont
  {Battaglia}}, \bibinfo {author} {\bibfnamefont {J.~B.}\ \bibnamefont
  {Hamrick}}, \bibinfo {author} {\bibfnamefont {V.}~\bibnamefont {Bapst}},
  \bibinfo {author} {\bibfnamefont {A.}~\bibnamefont {Sanchez-Gonzalez}},
  \bibinfo {author} {\bibfnamefont {V.}~\bibnamefont {Zambaldi}}, \bibinfo
  {author} {\bibfnamefont {M.}~\bibnamefont {Malinowski}}, \bibinfo {author}
  {\bibfnamefont {A.}~\bibnamefont {Tacchetti}}, \bibinfo {author}
  {\bibfnamefont {D.}~\bibnamefont {Raposo}}, \bibinfo {author} {\bibfnamefont
  {A.}~\bibnamefont {Santoro}}, \bibinfo {author} {\bibfnamefont
  {R.}~\bibnamefont {Faulkner}}, \emph {et~al.},\ }\bibfield  {title} {\bibinfo
  {title} {Relational inductive biases, deep learning, and graph networks},\
  }\href@noop {} {\bibfield  {journal} {\bibinfo  {journal} {arXiv preprint
  arXiv:1806.01261}\ } (\bibinfo {year} {2018})}\BibitemShut {NoStop}%
\bibitem [{\citenamefont {Bronstein}\ \emph {et~al.}(2017)\citenamefont
  {Bronstein}, \citenamefont {Bruna}, \citenamefont {LeCun}, \citenamefont
  {Szlam},\ and\ \citenamefont {Vandergheynst}}]{bronstein2017geometric}%
  \BibitemOpen
  \bibfield  {author} {\bibinfo {author} {\bibfnamefont {M.~M.}\ \bibnamefont
  {Bronstein}}, \bibinfo {author} {\bibfnamefont {J.}~\bibnamefont {Bruna}},
  \bibinfo {author} {\bibfnamefont {Y.}~\bibnamefont {LeCun}}, \bibinfo
  {author} {\bibfnamefont {A.}~\bibnamefont {Szlam}},\ and\ \bibinfo {author}
  {\bibfnamefont {P.}~\bibnamefont {Vandergheynst}},\ }\bibfield  {title}
  {\bibinfo {title} {Geometric deep learning: going beyond euclidean data},\
  }\href@noop {} {\bibfield  {journal} {\bibinfo  {journal} {IEEE Signal
  Processing Magazine}\ }\textbf {\bibinfo {volume} {34}},\ \bibinfo {pages}
  {18} (\bibinfo {year} {2017})}\BibitemShut {NoStop}%
\bibitem [{\citenamefont {Zhou}\ \emph {et~al.}(2020)\citenamefont {Zhou},
  \citenamefont {Cui}, \citenamefont {Hu}, \citenamefont {Zhang}, \citenamefont
  {Yang}, \citenamefont {Liu}, \citenamefont {Wang}, \citenamefont {Li},\ and\
  \citenamefont {Sun}}]{zhou2020graph}%
  \BibitemOpen
  \bibfield  {author} {\bibinfo {author} {\bibfnamefont {J.}~\bibnamefont
  {Zhou}}, \bibinfo {author} {\bibfnamefont {G.}~\bibnamefont {Cui}}, \bibinfo
  {author} {\bibfnamefont {S.}~\bibnamefont {Hu}}, \bibinfo {author}
  {\bibfnamefont {Z.}~\bibnamefont {Zhang}}, \bibinfo {author} {\bibfnamefont
  {C.}~\bibnamefont {Yang}}, \bibinfo {author} {\bibfnamefont {Z.}~\bibnamefont
  {Liu}}, \bibinfo {author} {\bibfnamefont {L.}~\bibnamefont {Wang}}, \bibinfo
  {author} {\bibfnamefont {C.}~\bibnamefont {Li}},\ and\ \bibinfo {author}
  {\bibfnamefont {M.}~\bibnamefont {Sun}},\ }\bibfield  {title} {\bibinfo
  {title} {Graph neural networks: A review of methods and applications},\
  }\href@noop {} {\bibfield  {journal} {\bibinfo  {journal} {AI Open}\ }\textbf
  {\bibinfo {volume} {1}},\ \bibinfo {pages} {57} (\bibinfo {year}
  {2020})}\BibitemShut {NoStop}%
\bibitem [{\citenamefont {Frydrych}\ \emph {et~al.}(2021)\citenamefont
  {Frydrych}, \citenamefont {Karimi}, \citenamefont {Pecelerowicz},
  \citenamefont {Alvarez}, \citenamefont {Dominguez-Guti{\'e}rrez},
  \citenamefont {Rovaris},\ and\ \citenamefont
  {Papanikolaou}}]{frydrych2021materials}%
  \BibitemOpen
  \bibfield  {author} {\bibinfo {author} {\bibfnamefont {K.}~\bibnamefont
  {Frydrych}}, \bibinfo {author} {\bibfnamefont {K.}~\bibnamefont {Karimi}},
  \bibinfo {author} {\bibfnamefont {M.}~\bibnamefont {Pecelerowicz}}, \bibinfo
  {author} {\bibfnamefont {R.}~\bibnamefont {Alvarez}}, \bibinfo {author}
  {\bibfnamefont {F.~J.}\ \bibnamefont {Dominguez-Guti{\'e}rrez}}, \bibinfo
  {author} {\bibfnamefont {F.}~\bibnamefont {Rovaris}},\ and\ \bibinfo {author}
  {\bibfnamefont {S.}~\bibnamefont {Papanikolaou}},\ }\bibfield  {title}
  {\bibinfo {title} {Materials informatics for mechanical deformation: A review
  of applications and challenges},\ }\href@noop {} {\bibfield  {journal}
  {\bibinfo  {journal} {Materials}\ }\textbf {\bibinfo {volume} {14}},\
  \bibinfo {pages} {5764} (\bibinfo {year} {2021})}\BibitemShut {NoStop}%
\bibitem [{\citenamefont {Bapst}\ \emph {et~al.}(2020)\citenamefont {Bapst},
  \citenamefont {Keck}, \citenamefont {Grabska-Barwi{\'n}ska}, \citenamefont
  {Donner}, \citenamefont {Cubuk}, \citenamefont {Schoenholz}, \citenamefont
  {Obika}, \citenamefont {Nelson}, \citenamefont {Back}, \citenamefont
  {Hassabis} \emph {et~al.}}]{bapst2020unveiling}%
  \BibitemOpen
  \bibfield  {author} {\bibinfo {author} {\bibfnamefont {V.}~\bibnamefont
  {Bapst}}, \bibinfo {author} {\bibfnamefont {T.}~\bibnamefont {Keck}},
  \bibinfo {author} {\bibfnamefont {A.}~\bibnamefont {Grabska-Barwi{\'n}ska}},
  \bibinfo {author} {\bibfnamefont {C.}~\bibnamefont {Donner}}, \bibinfo
  {author} {\bibfnamefont {E.~D.}\ \bibnamefont {Cubuk}}, \bibinfo {author}
  {\bibfnamefont {S.~S.}\ \bibnamefont {Schoenholz}}, \bibinfo {author}
  {\bibfnamefont {A.}~\bibnamefont {Obika}}, \bibinfo {author} {\bibfnamefont
  {A.~W.}\ \bibnamefont {Nelson}}, \bibinfo {author} {\bibfnamefont
  {T.}~\bibnamefont {Back}}, \bibinfo {author} {\bibfnamefont {D.}~\bibnamefont
  {Hassabis}}, \emph {et~al.},\ }\bibfield  {title} {\bibinfo {title}
  {Unveiling the predictive power of static structure in glassy systems},\
  }\href@noop {} {\bibfield  {journal} {\bibinfo  {journal} {Nature Physics}\
  }\textbf {\bibinfo {volume} {16}},\ \bibinfo {pages} {448} (\bibinfo {year}
  {2020})}\BibitemShut {NoStop}%
\bibitem [{\citenamefont {Shiba}\ \emph {et~al.}(2022)\citenamefont {Shiba},
  \citenamefont {Hanai}, \citenamefont {Suzumura},\ and\ \citenamefont
  {Shimokawabe}}]{shiba2022unraveling}%
  \BibitemOpen
  \bibfield  {author} {\bibinfo {author} {\bibfnamefont {H.}~\bibnamefont
  {Shiba}}, \bibinfo {author} {\bibfnamefont {M.}~\bibnamefont {Hanai}},
  \bibinfo {author} {\bibfnamefont {T.}~\bibnamefont {Suzumura}},\ and\
  \bibinfo {author} {\bibfnamefont {T.}~\bibnamefont {Shimokawabe}},\
  }\bibfield  {title} {\bibinfo {title} {Unraveling intricate processes of
  glassy dynamics from static structure by machine learning relative motion},\
  }\href@noop {} {\bibfield  {journal} {\bibinfo  {journal} {arXiv preprint
  arXiv:2206.14024}\ } (\bibinfo {year} {2022})}\BibitemShut {NoStop}%
\bibitem [{\citenamefont {Xie}\ and\ \citenamefont
  {Grossman}(2018)}]{xie2018crystal}%
  \BibitemOpen
  \bibfield  {author} {\bibinfo {author} {\bibfnamefont {T.}~\bibnamefont
  {Xie}}\ and\ \bibinfo {author} {\bibfnamefont {J.~C.}\ \bibnamefont
  {Grossman}},\ }\bibfield  {title} {\bibinfo {title} {Crystal graph
  convolutional neural networks for an accurate and interpretable prediction of
  material properties},\ }\href@noop {} {\bibfield  {journal} {\bibinfo
  {journal} {Physical review letters}\ }\textbf {\bibinfo {volume} {120}},\
  \bibinfo {pages} {145301} (\bibinfo {year} {2018})}\BibitemShut {NoStop}%
\bibitem [{\citenamefont {Pagan}\ \emph {et~al.}(2022)\citenamefont {Pagan},
  \citenamefont {Pash}, \citenamefont {Benson},\ and\ \citenamefont
  {Kasemer}}]{pagan2022graph}%
  \BibitemOpen
  \bibfield  {author} {\bibinfo {author} {\bibfnamefont {D.~C.}\ \bibnamefont
  {Pagan}}, \bibinfo {author} {\bibfnamefont {C.~R.}\ \bibnamefont {Pash}},
  \bibinfo {author} {\bibfnamefont {A.~R.}\ \bibnamefont {Benson}},\ and\
  \bibinfo {author} {\bibfnamefont {M.~P.}\ \bibnamefont {Kasemer}},\
  }\bibfield  {title} {\bibinfo {title} {Graph neural network modeling of
  grain-scale anisotropic elastic behavior using simulated and measured
  microscale data},\ }\href@noop {} {\bibfield  {journal} {\bibinfo  {journal}
  {arXiv preprint arXiv:2205.06324}\ } (\bibinfo {year} {2022})}\BibitemShut
  {NoStop}%
\bibitem [{\citenamefont {Nix}\ and\ \citenamefont
  {Gao}(1998)}]{nix1998indentation}%
  \BibitemOpen
  \bibfield  {author} {\bibinfo {author} {\bibfnamefont {W.~D.}\ \bibnamefont
  {Nix}}\ and\ \bibinfo {author} {\bibfnamefont {H.}~\bibnamefont {Gao}},\
  }\bibfield  {title} {\bibinfo {title} {Indentation size effects in
  crystalline materials: a law for strain gradient plasticity},\ }\href@noop {}
  {\bibfield  {journal} {\bibinfo  {journal} {Journal of the Mechanics and
  Physics of Solids}\ }\textbf {\bibinfo {volume} {46}},\ \bibinfo {pages}
  {411} (\bibinfo {year} {1998})}\BibitemShut {NoStop}%
\bibitem [{\citenamefont {Bolin}\ \emph {et~al.}(2019)\citenamefont {Bolin},
  \citenamefont {Yavas}, \citenamefont {Song}, \citenamefont {Hemker},\ and\
  \citenamefont {Papanikolaou}}]{bolin2019bending}%
  \BibitemOpen
  \bibfield  {author} {\bibinfo {author} {\bibfnamefont {R.}~\bibnamefont
  {Bolin}}, \bibinfo {author} {\bibfnamefont {H.}~\bibnamefont {Yavas}},
  \bibinfo {author} {\bibfnamefont {H.}~\bibnamefont {Song}}, \bibinfo {author}
  {\bibfnamefont {K.~J.}\ \bibnamefont {Hemker}},\ and\ \bibinfo {author}
  {\bibfnamefont {S.}~\bibnamefont {Papanikolaou}},\ }\bibfield  {title}
  {\bibinfo {title} {Bending nanoindentation and plasticity noise in fcc single
  and polycrystals},\ }\href@noop {} {\bibfield  {journal} {\bibinfo  {journal}
  {Crystals}\ }\textbf {\bibinfo {volume} {9}},\ \bibinfo {pages} {652}
  (\bibinfo {year} {2019})}\BibitemShut {NoStop}%
\bibitem [{\citenamefont {Durst}\ \emph {et~al.}(2006)\citenamefont {Durst},
  \citenamefont {Backes}, \citenamefont {Franke},\ and\ \citenamefont
  {G{\"o}ken}}]{durst2006indentation}%
  \BibitemOpen
  \bibfield  {author} {\bibinfo {author} {\bibfnamefont {K.}~\bibnamefont
  {Durst}}, \bibinfo {author} {\bibfnamefont {B.}~\bibnamefont {Backes}},
  \bibinfo {author} {\bibfnamefont {O.}~\bibnamefont {Franke}},\ and\ \bibinfo
  {author} {\bibfnamefont {M.}~\bibnamefont {G{\"o}ken}},\ }\bibfield  {title}
  {\bibinfo {title} {Indentation size effect in metallic materials: Modeling
  strength from pop-in to macroscopic hardness using geometrically necessary
  dislocations},\ }\href@noop {} {\bibfield  {journal} {\bibinfo  {journal}
  {Acta Materialia}\ }\textbf {\bibinfo {volume} {54}},\ \bibinfo {pages}
  {2547} (\bibinfo {year} {2006})}\BibitemShut {NoStop}%
\bibitem [{\citenamefont {Papanikolaou}\ \emph {et~al.}(2017)\citenamefont
  {Papanikolaou}, \citenamefont {Cui},\ and\ \citenamefont
  {Ghoniem}}]{papanikolaou2017avalanches}%
  \BibitemOpen
  \bibfield  {author} {\bibinfo {author} {\bibfnamefont {S.}~\bibnamefont
  {Papanikolaou}}, \bibinfo {author} {\bibfnamefont {Y.}~\bibnamefont {Cui}},\
  and\ \bibinfo {author} {\bibfnamefont {N.}~\bibnamefont {Ghoniem}},\
  }\bibfield  {title} {\bibinfo {title} {Avalanches and plastic flow in crystal
  plasticity: an overview},\ }\href@noop {} {\bibfield  {journal} {\bibinfo
  {journal} {Modelling and Simulation in Materials Science and Engineering}\
  }\textbf {\bibinfo {volume} {26}},\ \bibinfo {pages} {013001} (\bibinfo
  {year} {2017})}\BibitemShut {NoStop}%
\bibitem [{\citenamefont {Ruiz-Moreno}\ and\ \citenamefont
  {H{\"a}hner}(2018)}]{tem-1}%
  \BibitemOpen
  \bibfield  {author} {\bibinfo {author} {\bibfnamefont {A.}~\bibnamefont
  {Ruiz-Moreno}}\ and\ \bibinfo {author} {\bibfnamefont {P.}~\bibnamefont
  {H{\"a}hner}},\ }\bibfield  {title} {\bibinfo {title} {Indentation size
  effects of ferritic/martensitic steels: A comparative experimental and
  modelling study},\ }\href@noop {} {\bibfield  {journal} {\bibinfo  {journal}
  {Materials \& Design}\ }\textbf {\bibinfo {volume} {145}},\ \bibinfo {pages}
  {168} (\bibinfo {year} {2018})}\BibitemShut {NoStop}%
\bibitem [{\citenamefont {Kurpaska}\ \emph {et~al.}(2022)\citenamefont
  {Kurpaska}, \citenamefont {Dominguez-Gutierrez}, \citenamefont {Zhang},
  \citenamefont {Mulewska}, \citenamefont {Bei}, \citenamefont {Weber},
  \citenamefont {Kosi{\'n}ska}, \citenamefont {Chrominski}, \citenamefont
  {Jozwik}, \citenamefont {Alvarez-Donado} \emph {et~al.}}]{tem-2}%
  \BibitemOpen
  \bibfield  {author} {\bibinfo {author} {\bibfnamefont {L.}~\bibnamefont
  {Kurpaska}}, \bibinfo {author} {\bibfnamefont {F.}~\bibnamefont
  {Dominguez-Gutierrez}}, \bibinfo {author} {\bibfnamefont {Y.}~\bibnamefont
  {Zhang}}, \bibinfo {author} {\bibfnamefont {K.}~\bibnamefont {Mulewska}},
  \bibinfo {author} {\bibfnamefont {H.}~\bibnamefont {Bei}}, \bibinfo {author}
  {\bibfnamefont {W.}~\bibnamefont {Weber}}, \bibinfo {author} {\bibfnamefont
  {A.}~\bibnamefont {Kosi{\'n}ska}}, \bibinfo {author} {\bibfnamefont
  {W.}~\bibnamefont {Chrominski}}, \bibinfo {author} {\bibfnamefont
  {I.}~\bibnamefont {Jozwik}}, \bibinfo {author} {\bibfnamefont
  {R.}~\bibnamefont {Alvarez-Donado}}, \emph {et~al.},\ }\bibfield  {title}
  {\bibinfo {title} {Effects of fe atoms on hardening of a nickel matrix:
  Nanoindentation experiments and atom-scale numerical modeling},\ }\href@noop
  {} {\bibfield  {journal} {\bibinfo  {journal} {Materials \& Design}\ }\textbf
  {\bibinfo {volume} {217}},\ \bibinfo {pages} {110639} (\bibinfo {year}
  {2022})}\BibitemShut {NoStop}%
\bibitem [{\citenamefont {Gouldstone}\ \emph {et~al.}(2007)\citenamefont
  {Gouldstone}, \citenamefont {Chollacoop}, \citenamefont {Dao}, \citenamefont
  {Li}, \citenamefont {Minor},\ and\ \citenamefont {Shen}}]{tem-3}%
  \BibitemOpen
  \bibfield  {author} {\bibinfo {author} {\bibfnamefont {A.}~\bibnamefont
  {Gouldstone}}, \bibinfo {author} {\bibfnamefont {N.}~\bibnamefont
  {Chollacoop}}, \bibinfo {author} {\bibfnamefont {M.}~\bibnamefont {Dao}},
  \bibinfo {author} {\bibfnamefont {J.}~\bibnamefont {Li}}, \bibinfo {author}
  {\bibfnamefont {A.~M.}\ \bibnamefont {Minor}},\ and\ \bibinfo {author}
  {\bibfnamefont {Y.-L.}\ \bibnamefont {Shen}},\ }\bibfield  {title} {\bibinfo
  {title} {Indentation across size scales and disciplines: Recent developments
  in experimentation and modeling},\ }\href@noop {} {\bibfield  {journal}
  {\bibinfo  {journal} {Acta Materialia}\ }\textbf {\bibinfo {volume} {55}},\
  \bibinfo {pages} {4015} (\bibinfo {year} {2007})}\BibitemShut {NoStop}%
\bibitem [{\citenamefont {Kossman}\ and\ \citenamefont
  {Bigerelle}(2021)}]{kossman2021pop}%
  \BibitemOpen
  \bibfield  {author} {\bibinfo {author} {\bibfnamefont {S.}~\bibnamefont
  {Kossman}}\ and\ \bibinfo {author} {\bibfnamefont {M.}~\bibnamefont
  {Bigerelle}},\ }\bibfield  {title} {\bibinfo {title} {Pop-in identification
  in nanoindentation curves with deep learning algorithms},\ }\href@noop {}
  {\bibfield  {journal} {\bibinfo  {journal} {Materials}\ }\textbf {\bibinfo
  {volume} {14}},\ \bibinfo {pages} {7027} (\bibinfo {year}
  {2021})}\BibitemShut {NoStop}%
\bibitem [{\citenamefont {Dom{\'\i}nguez-Gut{\'\i}errez}\ \emph
  {et~al.}(2022)\citenamefont {Dom{\'\i}nguez-Gut{\'\i}errez}, \citenamefont
  {Mulewska}, \citenamefont {Ustrzycka}, \citenamefont {Alvarez-Donado},
  \citenamefont {Kos{\'\i}nska}, \citenamefont {Huo}, \citenamefont {Kurpaska},
  \citenamefont {Jozwik}, \citenamefont {Papanikolaou},\ and\ \citenamefont
  {Alava}}]{dominguez2022mechanisms}%
  \BibitemOpen
  \bibfield  {author} {\bibinfo {author} {\bibfnamefont {F.~J.}\ \bibnamefont
  {Dom{\'\i}nguez-Gut{\'\i}errez}}, \bibinfo {author} {\bibfnamefont
  {K.}~\bibnamefont {Mulewska}}, \bibinfo {author} {\bibfnamefont
  {A.}~\bibnamefont {Ustrzycka}}, \bibinfo {author} {\bibfnamefont
  {R.}~\bibnamefont {Alvarez-Donado}}, \bibinfo {author} {\bibfnamefont
  {A.}~\bibnamefont {Kos{\'\i}nska}}, \bibinfo {author} {\bibfnamefont
  {W.}~\bibnamefont {Huo}}, \bibinfo {author} {\bibfnamefont {L.}~\bibnamefont
  {Kurpaska}}, \bibinfo {author} {\bibfnamefont {I.}~\bibnamefont {Jozwik}},
  \bibinfo {author} {\bibfnamefont {S.}~\bibnamefont {Papanikolaou}},\ and\
  \bibinfo {author} {\bibfnamefont {M.}~\bibnamefont {Alava}},\ }\bibfield
  {title} {\bibinfo {title} {Mechanisms of strength and hardening in austenitic
  stainless 310s steel: Nanoindentation experiments and multiscale modeling},\
  }\href@noop {} {\bibfield  {journal} {\bibinfo  {journal} {arXiv preprint
  arXiv:2205.03050}\ } (\bibinfo {year} {2022})}\BibitemShut {NoStop}%
\bibitem [{\citenamefont {Bachmann}\ \emph {et~al.}(2010)\citenamefont
  {Bachmann}, \citenamefont {Hielscher}, \citenamefont {Jupp}, \citenamefont
  {Pantleon}, \citenamefont {Schaeben},\ and\ \citenamefont
  {Wegert}}]{bachmann2010inferential}%
  \BibitemOpen
  \bibfield  {author} {\bibinfo {author} {\bibfnamefont {F.}~\bibnamefont
  {Bachmann}}, \bibinfo {author} {\bibfnamefont {R.}~\bibnamefont {Hielscher}},
  \bibinfo {author} {\bibfnamefont {P.~E.}\ \bibnamefont {Jupp}}, \bibinfo
  {author} {\bibfnamefont {W.}~\bibnamefont {Pantleon}}, \bibinfo {author}
  {\bibfnamefont {H.}~\bibnamefont {Schaeben}},\ and\ \bibinfo {author}
  {\bibfnamefont {E.}~\bibnamefont {Wegert}},\ }\bibfield  {title} {\bibinfo
  {title} {Inferential statistics of electron backscatter diffraction data from
  within individual crystalline grains},\ }\href@noop {} {\bibfield  {journal}
  {\bibinfo  {journal} {Journal of Applied Crystallography}\ }\textbf {\bibinfo
  {volume} {43}},\ \bibinfo {pages} {1338} (\bibinfo {year}
  {2010})}\BibitemShut {NoStop}%
\bibitem [{\citenamefont {Oliver}\ and\ \citenamefont
  {Pharr}(1992)}]{oliver1992improved}%
  \BibitemOpen
  \bibfield  {author} {\bibinfo {author} {\bibfnamefont {W.~C.}\ \bibnamefont
  {Oliver}}\ and\ \bibinfo {author} {\bibfnamefont {G.~M.}\ \bibnamefont
  {Pharr}},\ }\bibfield  {title} {\bibinfo {title} {An improved technique for
  determining hardness and elastic modulus using load and displacement sensing
  indentation experiments},\ }\href@noop {} {\bibfield  {journal} {\bibinfo
  {journal} {Journal of materials research}\ }\textbf {\bibinfo {volume} {7}},\
  \bibinfo {pages} {1564} (\bibinfo {year} {1992})}\BibitemShut {NoStop}%
\end{thebibliography}%

\newpage
\renewcommand{\thefigure}{A\arabic{figure}}
\renewcommand{\thesection}{A~\Roman{section}}
\setcounter{figure}{0}    
\setcounter{section}{0}    
\section*{Appendix}
\begin{figure*}
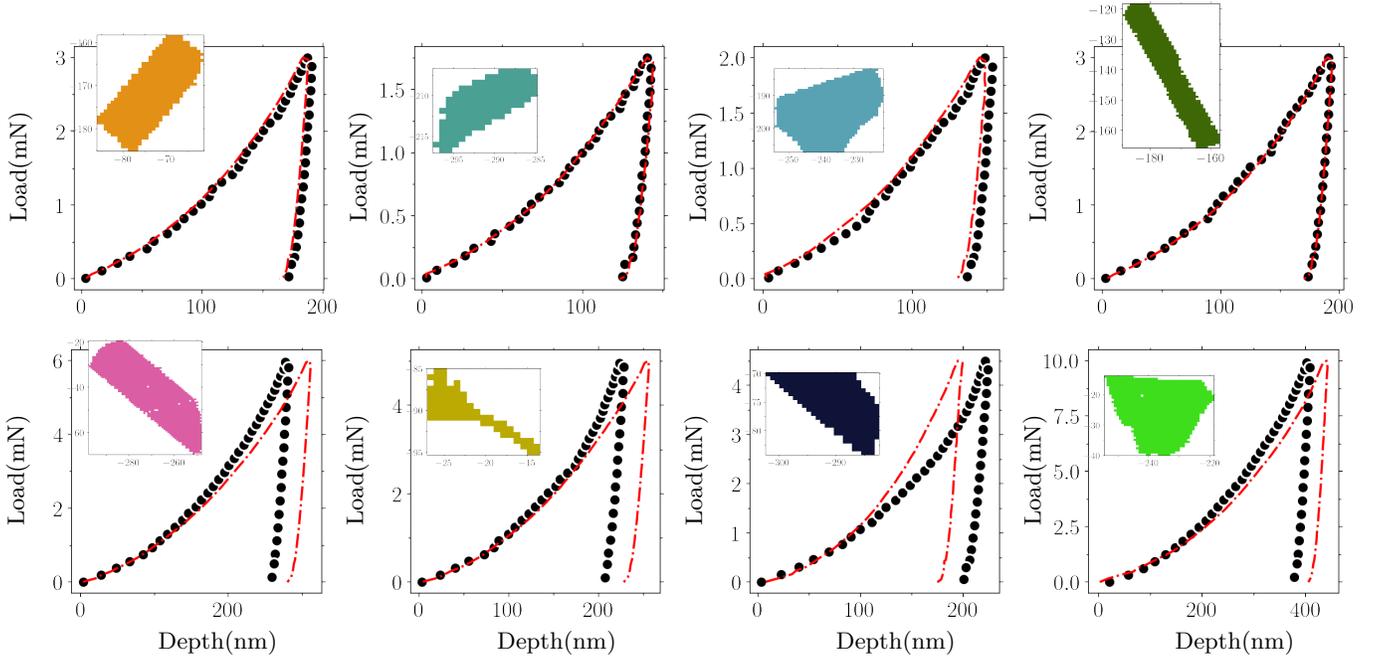

    \foreach \n in {0,...,3}{
    \begin{overpic}[width=0.24\textwidth]{Figs/prediction\n_10percent_loMSE.png}
        \put(14,50){\includegraphics[width=0.1\textwidth]{Figs/grain_indx\n_10percent_loMSE.png}}
        \Labelxy{-5}{32}{90}{{Load(mN)}}
    \end{overpic}
    }
    \foreach \n in {0,...,3}{
    \begin{overpic}[width=0.24\textwidth]{Figs/prediction\n_10percent_hiMSE.png}
        \put(14,50){\includegraphics[width=0.1\textwidth]{Figs/grain_indx\n_10percent_hiMSE.png}}
        \Labelxy{42}{-6}{0}{Depth(nm)}
        \Labelxy{-5}{32}{90}{{Load(mN)}}
    \end{overpic}
    }
    \caption{Actual (symbols) and predicted (dashdotted curve) load-depth curves associated with a subset of the indented grains (the insets) within the test set. The first and second rows indicate bottom $5\%$ (good predictions) and top $5\%$ (poor predictions) in mean-squared errors.}
    \label{fig:prediction_test}
\end{figure*}

In this appendix, we present the actual and predicted load-depth curves associated with a subset of the indented grains within the testing sets in Fig.~\ref{fig:prediction_test} and training sets in Fig.~\ref{fig:prediction_train}. The first and second rows indicate bottom $5\%$ (good predictions) and top $5\%$ (poor predictions) in mean-squared errors.

\begin{figure*}
    \foreach \n in {0,...,3}{
    \begin{overpic}[width=0.24\textwidth]{Figs/prediction\n_10percent_train_loMSE.png}
        \put(14,50){\includegraphics[width=0.1\textwidth]{Figs/grain_indx\n_10percent_train_loMSE.png}}
        \Labelxy{-5}{32}{90}{{Load(mN)}}
    \end{overpic}
    }
    \foreach \n in {0,...,3}{
    \begin{overpic}[width=0.24\textwidth]{Figs/prediction\n_10percent_train_hiMSE.png}
        \put(14,50){\includegraphics[width=0.1\textwidth]{Figs/grain_indx\n_10percent_train_hiMSE.png}}
        \Labelxy{42}{-6}{0}{Depth(nm)}
        \Labelxy{-5}{32}{90}{{Load(mN)}}
    \end{overpic}
    }
    \caption{Actual (symbols) and predicted (dashdotted curve) load-depth curves associated with a subset of the indented grains (insets) within the training sets. The first and second rows indicate bottom $5\%$ (good predictions) and top $5\%$ (poor predictions) in mean-squared errors.}
    \label{fig:prediction_train}
\end{figure*}

\end{document}


\title{Yielding in multi-component metallic glasses: Universal signatures of elastic modulus heterogeneities}

\author{Kamran Karimi$^1$}
\author{Mikko J. Alava$^{1,2}$}
\author{Stefanos Papanikolaou$^1$}
\affiliation{%
 $^1$ NOMATEN Centre of Excellence, National Center for Nuclear Research, ul. A. Sołtana 7, 05-400 Swierk/Otwock, Poland\\
 $^{2}$ Aalto University, Department of Applied Physics, PO Box 11000, 00076 Aalto, Espoo, Finland
}%

\maketitle

\section*{Supplementary Materials}
Simple shear tests were performed on the $xy$ plane at a fixed strain rate $\dot\gamma=10^{-5} ~\text{ps}^{-1}$ and temperature $T=300$ K up to a prescribed strain $\gamma \le 0.2$.  
We subsequently applied six deformation modes $\epsilon^{(m)}$, $m=1...6$ in Cartesian directions $xyz$ described by the following global strain tensors 
\begin{eqnarray}
\bm{\epsilon}^{(1)}&=&\epsilon_{xx}\delta_{\alpha x}\delta_{\beta x}, \nonumber \\ 
\bm{\epsilon}^{(2)}&=&\epsilon_{yy}\delta_{\alpha y}\delta_{\beta y}, \nonumber\\
\bm{\epsilon}^{(3)}&=&\epsilon_{zz}\delta_{\alpha z}\delta_{\beta z}, \nonumber\\
\bm{\epsilon}^{(4)}&=&\epsilon_{xy}(\delta_{\alpha x}\delta_{\beta y}+\delta_{\alpha y}\delta_{\beta x}), \nonumber\\
\bm{\epsilon}^{(5)}&=&\epsilon_{yz}(\delta_{\alpha y}\delta_{\beta z}+\delta_{\alpha z}\delta_{\beta y}), \nonumber\\
\bm{\epsilon}^{(6)}&=&\epsilon_{xz}(\delta_{\alpha x}\delta_{\beta z}+\delta_{\alpha z}\delta_{\beta x}).
\end{eqnarray}
Here $\delta_{\alpha\beta}$ is a Kronecker delta with $\alpha$ and $\beta$ denoting Cartesian indices.
The strain magnitudes are set to $\epsilon_{xx}=\epsilon_{yy}=\epsilon_{zz}=2\epsilon_{xy}=2\epsilon_{yxz}=2\epsilon_{xz}=10^{-2}$ in strain units.
We made use of (deformation-induced) stress differences $\Delta\sigma^{(m)}_{i,\alpha\beta}$, $m=1...6$ in atoms $i=1...N$ to evaluate the corresponding elasticity tensor
\begin{eqnarray}
c^i_{\alpha\beta xx}&=&\Delta\sigma^{(1)}_{i,\alpha\beta}/\epsilon_{xx}, \nonumber \\ 
c^i_{\alpha\beta yy}&=&\Delta\sigma^{(2)}_{i,\alpha\beta}/\epsilon_{yy}, \nonumber \\ 
c^i_{\alpha\beta zz}&=&\Delta\sigma^{(3)}_{i,\alpha\beta}/\epsilon_{zz}, \nonumber \\ 
c^i_{\alpha\beta xy}&=&\Delta\sigma^{(4)}_{i,\alpha\beta}/2\epsilon_{xy}, \nonumber \\ 
c^i_{\alpha\beta yz}&=&\Delta\sigma^{(5)}_{i,\alpha\beta}/2\epsilon_{yz}, \nonumber \\ 
c^i_{\alpha\beta xz}&=&\Delta\sigma^{(6)}_{i,\alpha\beta}/2\epsilon_{xz}.
\end{eqnarray}
It was also checked that the obtained elastic constants are almost insensitive to our choice of the above strain steps (i.e. the so-called \emph{linear} regime).

Statistics of clusters with $\mu=c^i_{xyxy}<0$ associated with \glzero, \glone, \gltwo, \glthree, \glfour, \glfive metallic glasses are illustrated in Fig.~\ref{fig:statistics_gamma} and \ref{fig:statistics_p}.
The fraction of unstable sites $p$ appears to evolve in a non-monotonic fashion, as shown in the leftmost panels of Fig.~\ref{fig:statistics_gamma}(a-f), with a peak value $p_\text{max}$ that almost coincides with that of the (bulk) shear stress $\sigma_{xy}$ (at $\gamma_{xy}\simeq 0.1$).
Overall, the evolution of $p$ with $\gamma_{xy}$ near $p_\text{max}$ features similar trends across all the compositions.
The center and right panels of Fig.~\ref{fig:statistics_gamma} show the mean cluster size $S$ and correlation length $\xi$ as a function of applied strain $\gamma_{xy}$ associated with different compositions.
Both observables tend to grow on approach to failure revealing slight compositional dependence.
This is further demonstrated in the first and third columns of Fig.~\ref{fig:statistics_p}(a-f) with the mean size and correlation length plotted as a function of $p$.
The rescaled data in the second and fourth columns of Fig.~\ref{fig:statistics_p}(a-f) feature a power-law scaling universally described by $S\propto (p_\text{max}-p)^{-\gamma}$ and $\xi\propto (p_{\text{max}}-p)^{-\nu}$ near \pmax.
Here, we note that every panel in the second and fourth columns shows two decades in $S$ and slightly over one decade in $\xi$.
The estimated critical exponents $\gamma$ and $\nu$ are reported in Table~$1$.

\begin{figure*}
    \foreach \n in {0,...,5}{
    \begin{overpic}[width=0.23\textwidth]{Figs/p_gamma_glass\n.png}
        \LabelFig{12}{87}{\glass{\n}}
        \ifnum\n=5{\Labelxy{50}{-3}{0}{$\gamma_{xy}$}} \fi
        \Labelxy{-8}{46}{90}{{$p$}}
    \end{overpic}
    \begin{overpic}[width=0.23\textwidth]{Figs/smean_gamma_glass\n.png}
        \ifnum\n=5{\Labelxy{50}{-3}{0}{$\gamma_{xy}$}} \fi
        \Labelxy{-8}{46}{90}{{$S\scriptstyle(\times 10^{3})$}}
    \end{overpic}
    \begin{overpic}[width=0.23\textwidth]{Figs/crltnl_gamma_glass\n.png}
        \ifnum\n=5{\Labelxy{50}{-3}{0}{$\gamma_{xy}$}} \fi
        \Labelxy{-8}{46}{90}{{$\xi$\tiny (\r{A})}}
    \end{overpic}
    
    \vspace{7pt}
    }
    \caption{Cluster statistics: probability $p$ ($1$st column), mean size $S$ ($2$nd column), and correlation length $\xi$ ($3$rd column) plotted against applied strain $\gamma_{xy}$ corresponding to the \glzero, \glone, \gltwo, \glthree, \glfour, and \glfive glasses. The (red) curves indicate binning averaged data.}
    \label{fig:statistics_gamma}
\end{figure*}

\begin{figure*}
    \foreach \n in {0,...,5}{
    \begin{overpic}[width=0.23\textwidth]{Figs/smean_p\n.png}
        \Labelxy{-8}{46}{90}{{$S\scriptstyle(\times 10^{3})$}}
        \LabelFig{15}{86}{\glass{\n}}
        \ifnum\n=5{\Labelxy{50}{-3}{0}{$p$}}\fi
    \end{overpic}
    \begin{overpic}[width=0.23\textwidth]{Figs/smean_p_rescaled\n.png}
        \Labelxy{-8}{46}{90}{{$S$}}
        \ifnum\n=5{\Labelxy{38}{-3}{0}{$1-p/p_\text{max}$}} \fi
        \ifnum\n=0 {
        \begin{tikzpicture}
          \coordinate (a) at (0,0);
            \node[white] at (a) {\tiny.};               %
            \drawSlope{2.4}{2.8}{0.3}{224}{darkred}{$\hspace{-2pt}\gamma$}
		\end{tikzpicture}} \else \fi
    \end{overpic}
    \begin{overpic}[width=0.23\textwidth]{Figs/crltnl_p\n.png}
        \Labelxy{-8}{46}{90}{{$\xi$\tiny (\r{A})}}
        \ifnum\n=5{\Labelxy{50}{-3}{0}{$p$}}\fi
    \end{overpic}
    \begin{overpic}[width=0.23\textwidth]{Figs/crltnl_p_rescaled\n.png}
        \Labelxy{-8}{46}{90}{{$\xi$\tiny (\r{A})}}
        \ifnum\n=5{\Labelxy{38}{-3}{0}{$1-p/p_\text{max}$}}\fi
        \ifnum\n=0 {
        \begin{tikzpicture}
          \coordinate (a) at (0,0);
            \node[white] at (a) {\tiny.};               %
            \drawSlope{1.3}{2.8}{0.3}{231}{darkred}{$\hspace{-2pt}\nu$}
		\end{tikzpicture}} \else \fi
    \end{overpic}
    
    \vspace{10pt}
    }
    \caption{Cluster statistics: mean size $S$ ($1$st and $2$nd columns) and correlation length $\xi$ ($3$rd and $4$th columns) plotted against probability $p$ and $1-p/p_\text{max}$ corresponding to the \glzero, \glone, \gltwo, \glthree, \glfour, and \glfive glasses. The (red) curves indicate binning averaged data. The dashdotted lines denote power laws $S\propto (p_{\text{max}}-p)^{-\gamma}$  and $\xi\propto (p_{\text{max}}-p)^{-\nu}$. The scaling exponents $\gamma$ and $\nu$ are reported in Table~1.}
    \label{fig:statistics_p}
\end{figure*}